\documentclass[journal=aamick, manuscript=article, layout=twocolumn]{achemso}
\usepackage{sansmath} 
\sansmath 
\usepackage[version=3]{mhchem} 
\usepackage{siunitx}

\usepackage[version=3]{mhchem} 

\usepackage{xcolor}

\author{Konrad Wilke}%
\affiliation{%
Department of Applied Physics and Science Education, Eindhoven University of Technology, 5600 MB, Eindhoven, The Netherlands }

\author{Mike Pols}
\affiliation{%
Department of Applied Physics and Science Education, Eindhoven University of Technology, 5600 MB, Eindhoven, The Netherlands }

\author{Titus S. van Erp}
\affiliation{%
Department of Chemistry and Biomedical Science,  Norwegian University of Science and Technology, 7941 Trondheim, Norway
}

\author{Geert Brocks}%
\affiliation{%
Department of Applied Physics and Science Education, Eindhoven University of Technology, 5600 MB, Eindhoven, The Netherlands; Computational Chemical
Physics, Faculty of Science and Technology and MESA+ Institute for Nanotechnology, University of Twente, Enschede
7500 AE, The Netherlands.}

\author{Shuxia Tao}
\email{s.x.tao@tue.nl}
\affiliation{%
Department of Applied Physics and Science Education, Eindhoven University of Technology, 5600 MB, Eindhoven, The Netherlands }

\date{\today}

\title{Anisotropic Defect Diffusion in Layered CsPbBr$_\mathrm{x}$I$_\mathrm{3-x}$ Perovskites} 

\keywords{Halide perovskites, defects, diffusion, migration, reactive force field, ion migration, strain engineering, ordering, stability, optoelectronics}

\begin{document}
\begin{abstract}
Mixed-halide perovskites offer a route to enhance phase stability and modify optoelectronic properties. Here, we use large-scale molecular dynamics simulations with a reactive force field to investigate defects in CsPbBr$_\mathrm{x}$I$_\mathrm{3-x}$ perovskites, focusing on how defect mobility can be controlled and the stability of the material improved by layered ordering of Br and I anions. Our results show that layered halide ordering induces strongly anisotropic defect diffusion: migration proceeds readily along the layers, whereas diffusion across them is strongly suppressed. For Cs defects, this anisotropy originates from directional lattice strain and the associated octahedral tilting, whereas halide migration is governed by an interplay between strain and preferential local halide bonding configurations.

\end{abstract}






\section{Introduction}
Metal halide perovskites (MHPs) with general formula ABX$_3$ (A = monovalent cation, B = metal, X = halide) have revolutionized optoelectronics due to their tunable band gaps\cite{Filip2014,Unger2017,Niebur2025}, high carrier mobilities\cite{Le_Corre2021,Herz2017,Kumar_Tailor2021}, and low-cost processability\cite{Snaith2013,Zhang2016,Soto-Montero2024}. Extensive research over the past two decades has resulted in power conversion efficiencies exceeding 26\% in photovoltaic applications\cite{Green2025}, yet challenges remain regarding structural stability\cite{Xiang2021,Jiang2020,Leijtens2015,Chen2021,Steele2020}.

All-inorganic halide perovskites, CsPbX$_3$ (X = Cl, Br, I) in particular, offer improved thermal stability relative to their organic-inorganic counterparts, as they do not suffer from the decomposition reactions that affect organic molecules such as methylammonium (MA).\cite{Chang2016,Yao2021} The 1.73 eV band gap of CsPbI$_3$\cite{Eperon2014} is relatively close to the optimal value of 1.34 eV required for the Shockley-Queisser limit for photovoltaic efficiency.\cite{Shockley1961} However, the photoactive black perovskite phase of CsPbI$_3$ is only metastable at room temperature, and susceptible to a phase transition to the more stable non-perovskite yellow $\delta$-phase.\cite{Jin2026,Stoumpos2015,Marronnier2018,Straus2020,Steele2022} 

The black phase can be stabilized by replacing some I by Br ions, thereby improving the size compatibility of the ions in the lattice, in accordance with the Goldschmidt tolerance factor.\cite{Goldschmidt1926,Nasstrom2020} Unfortunately, the band gap then also increases (with CsPbBr$_3$ having a band gap of 2.37 eV)\cite{Mannino2020}, making the material less suitable for single-junction solar cells.\cite{Chen2021, Li2019a} 

One strategy to alleviate this could be a layered halide ordering, which in the case of CsPbBr$_\mathrm{2}$I reduces the band gap by up to 0.2~eV compared to a random alloy.\cite{Deng2023} The main idea is that ordering halide ions of different size introduces anisotropy in the strain field.\cite{Teunissen2023} This limits the octahedral tilting of the PbX$_6$ octahedrons, which is predicted to reduce the band gap.\cite{Deng2023,Steele2021}

\citeauthor{Deng2023} recently demonstrated layered halide orderings in CsPbBr$_\mathrm{x}$I$_\mathrm{3-x}$ through solvent evaporation via the antisolvent protection method.\cite{Deng2023} Using the same method, \citeauthor{Lin2026} obtained ordered DMAPbBr$_\mathrm{x}$I$_\mathrm{3-x}$ (DMA = dimethylammonium) single crystals as precursors for layer-ordered CsPbBr$_\mathrm{x}$I$_\mathrm{3-x}$ thin films.\cite{Lin2026} Such ordering reduces the band gap and increases carrier mobility.\cite{Deng2023,Lin2026} 

Interestingly, besides the ability to tune optoelectronic properties, halide ordering can also increase the stability of CsPbBr$_\mathrm{x}$I$_\mathrm{3-x}$ perovskite phase beyond what would be expected on the basis of the Goldschmidt tolerance factor.\cite{Deng2023} A contributing factor could again be the anisotropic strain field. Indeed, a stabilizing influence of biaxial strain on CsPbI$_3$ and CsPbBr$_\mathrm{x}$I$_\mathrm{3-x}$ ($\mathrm{x} \le 0.1$) has been reported experimentally, and ascribed to a reduction of the energy difference between the metastable black phase and the yellow phase.\cite{Steele2019} 

In addition to the thermodynamic effects of strain, there may also be an effect on the kinetics of degradation. Lattice point defects, such as vacancies and interstitials, can play a pivotal role in destabilizing halide perovskite phases.\cite{Goldschmidt1926,Pols2021,Chen2021,Delugas2016a,Mattoni2024, Phung2020} For example, \citeauthor{Pols2021} demonstrated that iodine vacancies are highly mobile and act as degradation initiators.\cite{Pols2021} \citeauthor{Delugas2016a} observed anisotropic halide defect migration and suggested controlling it with strain and composition.\cite{Delugas2016a}

In this paper, we explore ordering the halide species in CsPbBr$_\mathrm{x}$I$_\mathrm{3-x}$ to steer the migration of point defects, and effectively block it in certain directions. We investigate the directional defect diffusion in CsPbBr$_\mathrm{x}$I$_\mathrm{3-x}$ perovskites by means of molecular dynamics (MD) simulations. We consider the motion of anion vacancies and interstitials of Br and I, but also that of Cs cation defects, which we show have a comparable mobility. By comparing layered and randomly mixed halide orderings, we disentangle the effects of chemical heterogeneity and strain-induced structural distortions on defect behavior. To isolate the influence of lattice strain alone, we apply controlled biaxial strain to pure $\alpha$-CsPbI$_3$ and $\alpha$-CsPbBr$_3$ structures. This approach allows us to systematically probe how both chemical composition and mechanical strain govern defect energetics and migration pathways and dynamics. Our study offers atomistic insight of strain engineering into defect-mediated transport in layered halide perovskites. It informs design strategies to optimize ionic and electronic properties for high-performance optoelectronic applications.

\section{Computational Methods}
\subsection{Structural Models}
\begin{figure*}[h!]
  \centering
  \begin{tabular}{ccc}
    \includegraphics[]{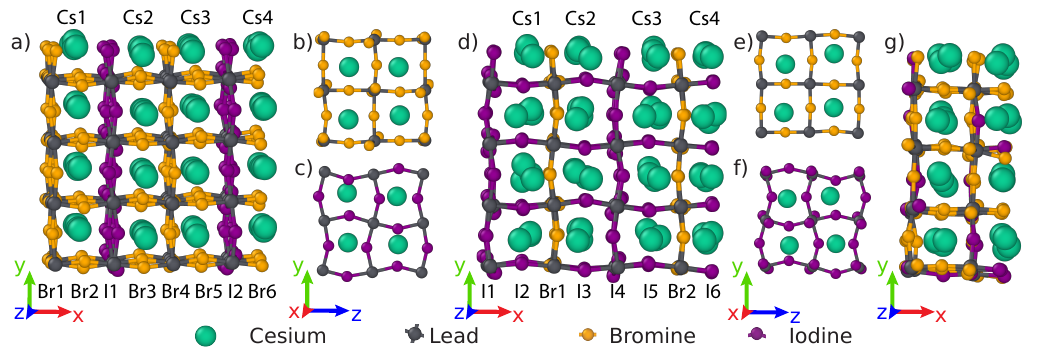}
  \end{tabular}
  \caption{a) Supercell of od-CsPbBr$_2$I with Br and I layers stacked along the $x$-direction; b), c) top views of Br and I layers. d), e), f), similar for od-CsPbBrI$_2$; g) ud-CsPbBr$_2$I.}
  \label{fig:structures}
\end{figure*}

To study the effects of layered halide ordering on defect migration and the structure of CsPbBr$_\mathrm{x}$I$_\mathrm{3-x}$, we focus on the ordered (od-) CsPbBr$_2$I compound (Fig.~\ref{fig:structures}a)-c)) synthesized in Ref.~\citenum{Deng2023}. For comparison, we also consider the iodine-rich, ordered counterpart od-CsPbBrI$_2$ (Fig.~\ref{fig:structures}d)-f)) and the corresponding structures with unordered (ud) halides: ud-CsPbBr$_2$I (Fig.~\ref{fig:structures}g)) and ud-CsPbBrI$_2$. In the od-structures, the majority halide ions fill all sites on every other Pb-X layer and the neighboring Cs-X layers, and the minority halide ions fill the remaining Pb-X layers (see Fig.~\ref{fig:structures}). We compare results obtained with these ordered structures with results obtained with random structures of the same composition. 

We construct $4\times4\times4$ supercells of the primitive cubic cell of the high temperature phases of CsPbI$_3$ and CsPbBr$_3$: $\alpha$-CsPbI$_3$ and $\alpha$-CsPbBr$_3$ (space group $\mathrm{Pm\bar{3}m}$) with 320 atoms. The ordered structures are constructed as described above; the randomly mixed structures are generated by randomly positioning I and Br on halide positions, maintaining the Br:I ratio in all space directions. The defects are introduced into the systems as single I or Br vacancies or interstitials, or as single Cs vacancies or interstitials, inside the supercell. It is important to note that charge states, as presented by \citeauthor{Tyagi2025}\cite{Tyagi2025}, cannot be easily accounted for in ReaxFF, as the training set includes only neutral charge states. However, the force field uses a dynamic charge-equilibration scheme to account for polarization effects.\cite{Senftle2016}

\subsection{Molecular dynamics simulations}

We employ the ReaxFF parameterization of \citeauthor{Pols2024} in this work.\cite{Pols2024} It extends the previous ReaxFF parameterization for CsPbI$_3$\cite{Pols2021} to mixed-halide CsPbBr$_\mathrm{x}$I$_\mathrm{3-x}$ systems. It provides an accurate structural description of complex environments, including the octahedral tilts in mixed perovskite structures. The computational efficiency of ReaxFF makes it suitable for long-time simulations. Alternatively, machine-learned force fields (MLFFs) have also been applied to related compositions.\cite{Pols2023,Tyagi2025,Tyagi2026} Whereas the latter approach can be more accurate in principle, it allows for controlling the charge state of defects, for instance, it is also computationally more demanding. A detailed validation and efficiency benchmark of the chosen ReaxFF model is provided in the SI~Note:~4.
 
All molecular dynamics (MD) simulations were performed in \texttt{AMS2024}.\cite{Ruger_2024} The simulations were performed at 700~K, with the standard ReaxFF time step of 0.25~fs and an output frequency of 0.1~ps. Production runs were prepared with 50~ps of $NVT$ equilibration and a Berendsen thermostat\cite{Berendsen1984} with a damping constant of $\tau_T=100~\mathrm{fs}$, which was also applied in all further runs. The atomic positions were taken from the last frame of the respective equilibration. The production runs used a Nose-Hoover chain thermostat\cite{Martyna1992} with a chain length of 10. The production runs were carried out for 10~ns in the $NVT$ ensemble, with an output frequency of 0.5~ps.

The volume of the supercell was determined in a simulation of the defect-free system. For this simulation, the equilibration was performed for 50~ps within the $NpT$ ensemble using a Berendsen thermostat and barostat\cite{Berendsen1984} and damping constant $\tau_p=2500~\mathrm{fs}$. The box dimensions were set to the average of the last 25~ps of the $NpT$ equilibration.

Defect tracking was performed using a dynamic site-based analysis that follows the idea of the site-projection approach implemented in the \texttt{site-analysis} code~\cite{Morgan2026}. For each defect species, the simulation supercell is partitioned into discrete, space-filling, non-overlapping volumes (sites), constructed separately for each MD frame from the instantaneous positions of the framework Pb atoms. Mobile Cs and I/Br atoms are then assigned to a specific site at each MD frame by projection onto these volumes.

To track all defects, we used a dynamic Voronoi decomposition of the simulation cell. At every MD frame, the midpoint of each pair (halides) or octet (Cs) of nearest-neighbor Pb atoms was computed from their instantaneous positions, with these midpoints used as seeds for a standard (per-frame) Voronoi decomposition. In the non-defective structure, every site is occupied by exactly one Cs or halide atom. Therefore, we associate unoccupied sites with vacancies, and doubly occupied sites with interstitials.

The mean squared displacement (MSD) has been calculated with the \texttt{AMS MSD} tool,\cite{Ruger_2024} according to
\begin{equation}
  \mathrm{MSD}(t) = \langle N^{-1} \sum_{i}^{N} |\mathbf{r}_i(t) - \mathbf{r}_i(0)|^2 \rangle ,
\end{equation}
where $N$ is the number of particles of the migrating species, $\mathbf{r}_i(t)$ the position of particle $i$ at time $t$ and $\mathbf{r}_i(0)$ the initial position. The brackets indicate an average over all time origins with a time window of 5 ns. The diffusion coefficient $D$ is obtained through a linear fit of the MSD according to
\begin{equation}
  D = \frac{1}{2d} \frac{\mathrm{d}}{\mathrm{d}t} \mathrm{MSD}(t),
\end{equation}
where $d$ is the dimensionality of the diffusion. For the determination of directional diffusion coefficients, we calculate the MSD separately for each Cartesian direction and set $d=1$. 

All atomic structures shown in this work were generated using \texttt{OVITO} version 3.10.2.\cite{Stukowski2009}. Furthermore, we used its implementation of polyhedral template matching\cite{Larsen2016} to determine the PbX$_6$ octahedra.

\section{Results and discussion}

To systematically probe the influence of ordered halide layers on defect migration, we evaluated mass transport in od-CsPbBr$_2$I and od-CsPbBrI$_2$ and compared it to that in randomly mixed ud-CsPbBr$_\mathrm{x}$I$_\mathrm{3-x}$ ($\mathrm{x}=1$ or $2$) and in CsPbI$_3$ and CsPbBr$_3$. Each system contains a single point defect, denoted as V$_S$ for a vacancy or I$_S$ for an interstitial with $S=$ {I, Br, Cs}. We focus primarily on the atomistic migration behavior of the defect, using the Voronoi cell method to track its migration path. Subsequently, we link this to the macroscopic diffusion behavior of the corresponding ionic species.

We first consider Cs vacancies and interstitials and elucidate the effect of anisotropic strain introduced by the layered halide structure on their motion. Next we focus on halide vacancies and interstitials, establishing the role played by the different bonding strengths of Pb-Br and Pb-I bonds. We do not look at Pb point defects, as these are immobile at the temperature and time range considered here.\cite{Eames2015,Eperon2017}

\subsection{Cesium defects and strain}
Experimental and computational studies of defect motion in halide perovskites typically focus on halide defects. Although these are certainly important, A-site cation defects, Cs vacancies and interstitials in this case, are also mobile.\cite{Eperon2015a,Niemann2016,Hu2016} First-principles studies show A-site defects should be prevalent, and compound defects further complicate the defect landscape, motivating explicit study of Cs migration.\cite{Xue2022,Xue2023} Despite A-site defects being electronically benign, A-site ions stabilize the perovskite lattice, so their migration may trigger structural phase changes.  

\begin{figure}[h!]
    \centering
      \includegraphics[]{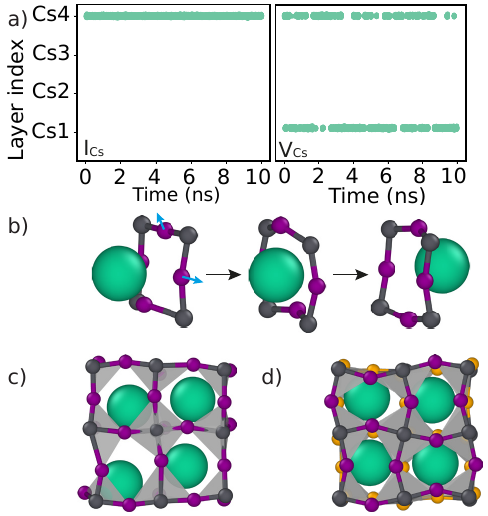} \\
    \caption{a) Positions of interstitial and vacancy, I$_\mathrm{Cs}$, V$_\mathrm{Cs}$, vs. simulation time in od-CsPbBr$_2$I (for position labels, see Fig. \ref{fig:structures}); b) migration event with gate opening in $\alpha$-CsPbI$_3$; snapshots of c) $\alpha$-CsPbI$_3$ and d) od-CsPbBr$_2$I structures.}   
    \label{fig:cs_jumps_across_x}
\end{figure}

Using the Voronoi-scheme we track the position over time of V$_\mathrm{Cs}$ and I$_\mathrm{Cs}$. The results are shown in Fig.~\ref{fig:cs_jumps_across_x}a). For V$_\mathrm{Cs}$ jumps to or across Pb-I layers are strongly suppressed, and motion predominantly occurswithin the Pb-Br layers. Diffusion of I$_\mathrm{Cs}$ is even more anisotropic and strictly two-dimensional throughout the entire simulation.  

\begin{figure*}[h!]
  \centering
  \includegraphics[]{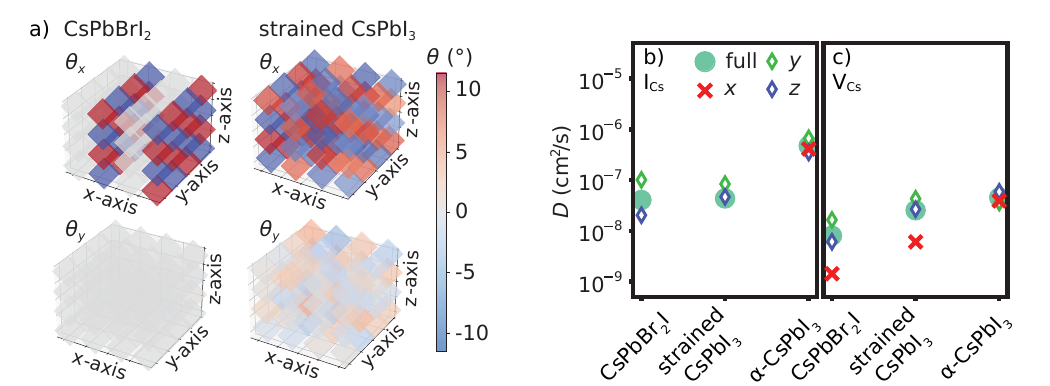}
  \caption{a) Time-averaged octahedral tilt angles with respect to $x$ and $y$ axes, $\theta_x$ and $\theta_y$, in CsPbBr$_2$I and strained CsPbI$_3$ (full tilt analysis in SI~Note~3); b), c) diffusion coefficients of Cs interstitial and vacancy, I$_\mathrm{Cs}$, V$_\mathrm{Cs}$, in CsPbBr$_2$I, strained CsPbI$_3$, and unstrained $\alpha$-CsPbI$_3$. Diffusion of I$_\mathrm{Cs}$ in od-CsPbBr$_2$I and strained CsPbI$_3$ was too slow to gather enough data for the MSD analysis.}
  \label{fig:cs_diffusion}
\end{figure*}

The migration of Cs through the lattice is conceptually simple. For both interstitial and vacancy, a migration event consists of a Cs species crossing one of the sides of the cuboid formed by the eight PbX$_{6}$ octahedrons surrounding it. Such a crossing is accompanied by a gate-opening rearrangement of the surrounding octahedrons, as shown in Fig.~\ref{fig:cs_jumps_across_x}b). In $\alpha$-CsPbBr$_3$ and $\alpha$-CsPbI$_3$, these rearrangements occur along all three Cartesian axes with equal probability at high temperatures. Layered MPHs show more structural rigidity in certain directions, as illustrated by comparing the octahedral tilts and window shapes of $\alpha$-CsPbI$_3$ (Fig.~\ref{fig:cs_jumps_across_x}c) with those in CsPbBr$_2$I (Fig.~\ref{fig:cs_jumps_across_x}d).

We propose that the anisotropic diffusion of Cs defects is caused by the strain induced by linking the differently sized PbI$_4$X$_2$ and PbBr$_4$X$_2$ octahedra, which forces the PbI$_4$X$_2$ octahedra to tilt, as shown in Fig.~\ref{fig:cs_jumps_across_x}d). This locks Pb-I-Pb angles and suppresses the gate-opening motion required for a Cs atom to jump to the adjacent lattice site. 

To quantify this strain-driven picture, we monitor the octahedral tilting angles relative to the cartesian axes and calculate their time averages ($\theta_x$, $\theta_y$ and $\theta_z$) along the MD trajectories. The octahedra in the Pb-I layers of the CsPbBr$_2$I lattice are tilted with respect to the normal to those layers, whereas in both other dimensions the octahedral tilt is negligible ($|\theta_x|\approx 12^\mathrm{o}$, $\theta_y \approx 0$, $\theta_z \approx 0 $). The tilts of neighboring octahedra have alternating signs, arranged in a checkerboard-like pattern, cf. Fig.~\ref{fig:cs_diffusion}a). The octahedra in the Pb-Br layers have time-averaged tilting angles close to zero.  

To isolate the effect of strain from chemical composition, we apply biaxial strain to pristine $\alpha$-CsPbI$_3$ and $\alpha$-CsPbBr$_3$ structures to replicate the octahedral tilt patterns observed in layered mixed-halide perovskites. In particular, we fix the lattice constant of $\alpha$-CsPbI$_3$ in the $x$-direction, and apply compressive strain by reducing the lattice constants in the $y$- and $z$-directions by 4\%. In an experimental setting, \citeauthor{Steele2019} applied biaxial strain by clamping a CsPbI$_3$ substrate.\cite{Steele2019} They could achieve a strain of 1.65\% and reported an increase in stability.\cite{Steele2019}

On an atomic level, this results in a checkerboard octahedral tilting pattern with $\theta_x$ similar to that observed in CsPbBr$_2$I, see Fig.~\ref{fig:cs_diffusion}a). There is some tilting in the $y$- and $z$-direction, but since $|\theta_y| \approx |\theta_z| = 2.1^\mathrm{o} \ll |\theta_x| = 9.4^\mathrm{o}$, this structure should reasonably represent that of the Pb-I layers in CsPbBr$_2$I. In a similar way, tensile strain is applied to the $\alpha$-CsPbBr$_3$ lattice to confirm that no octahedral tilting is introduced in Pb-Br layers. The exact procedure and tests can be found in the SI~Note~3.3.

We have determined the diffusion coefficients at $T=700$K of Cs interstitials and vacancies in $x$-, $y$- and $z$-directions in strained and $\alpha$-CsPbI$_3$ and compare them with those in CsPbBr$_2$I; the results are shown in Figures~\ref{fig:cs_diffusion}b) and c). As one expects, $\alpha$-CsPbI$_3$, which has a cubic structure, shows isotropic migration for Cs defects. 
In strained CsPbI$_3$, diffusion becomes anisotropic, and, what is more important, the directional diffusion coefficients become very similar to those in CsPbBr$_2$I. 

Both V$_\mathrm{Cs}$ and I$_\mathrm{Cs}$ show a pronounced preference for migration within the layers. In the case of Cs interstitials we even see a complete suppression of migration events in $x$-direction, as shown in Fig.~\ref{fig:cs_diffusion}b). 
Tab.~\ref{tab:cs_int_orientation} gives the residence times of interstitials being oriented along the $x$-axis. We observe that this is only the case in a small time fraction of the simulation, which limits the overall attempt rate of a jump in the $x$-direction. In other words, one expects Cs interstitials to be confined perpendicular to $x$ in $yz$ layers.
V$_\mathrm{Cs}$ is confined less strongly, but still the diffusion coefficients in $y$-, and $z$-directions are an order of magnitude larger than that in the $x$-direction, Fig.~\ref{fig:cs_diffusion}c). 

\begin{table}[h!]
\centering
\caption{The fraction of time  I$_\mathrm{Cs}$ is oriented along the $x$-axis and perpendicular to it during the simulation. Without structural bias, we expect: $\perp x :$~$\parallel x = 2:1$. }
\label{tab:cs_int_orientation}
\begin{tabular}{|l|c|c|}
\hline
Config & $\perp x$ (\%) & $\parallel x$ (\%) \\
\hline
\hline
$\alpha$-CsPbI$_3$  & 66.7 & 33.3 \\
od-CsPbBr$_2$I & 98.0 & 2.0 \\
strained CsPbI$_3$ & 93.4 & 6.6 \\
\hline
\end{tabular}
\end{table}

To explain the structural origin of the diffusion anisotropy, we note that the checkerboard-like tilting pattern of the PbX$_6$ octahedra, with $|\theta_x|\neq 0$ and $|\theta_y|\approx 0$, $|\theta_z|\approx 0$, locks specific Pb--I--Pb angles in the $yz$-plane (Fig.~\ref{fig:cs_jumps_across_x}d)). This constrains the gate-opening motion required for a Cs atom to jump to a neighboring lattice site along the $x$-direction (Fig.~\ref{fig:structures}). We expect this effect, and thus the resulting anisotropy, to become even more pronounced at lower temperatures because octahedral tilting in the iodine layers increases (see SI~Note~3.2), consistent with general trends in mixed-halide perovskites.\cite{Pols2024,Teunissen2023}

Furthermore, we studied the influence of tensile strain along the $y$- and $z$-axes of the $\alpha$-CsPbBr$_3$ lattice. Here, a slight reduction of octahedral tilts was accompanied by an approximately two times higher diffusion coefficients in the $x$-direction compared to $y$ and $z$. Concrete values can be found in the SI~Tab.~S1 in Note~1 and Fig.~S7 in Note~3.3.

\subsection{Halide defects and chemical composition}

Halide defects have been widely studied in MHPs, as they relatively easiy to form and highly mobile, and thus are thought to strongly influence ionic conductivity and structural and chemical stability.\cite{Pols2021,Tyagi2025,Pols2023,Xue2022,Xue2023} Similar to the Cs defects, we observe confinement to layers for both vacancies and interstitials. This effect appears to increase when extrapolating to lower temperatures, as evident from the activation energies reported in SI Note~4.3 (Tab.s~S8 and~S9).

However, the underlying reasons differ somewhat, as strain is not the only factor guiding halide point defect diffusion; the chemical difference between Br and I plays a distinct role. 

To isolate the role of halide composition from directional strain effects, we compare diffusion in od-CsPbBr$_2$I and od-CsPbBrI$_2$ structures with that in ud-CsPbBr$_2$I and ud-CsPbBrI$_2$. We only present the results for a removed and added Br here, as the results for I defects are qualitatively similar. Results for the latter can be found in the SI~Note~2. 

The basic reason for these similarities is that both interstitial and vacancy migration typically proceed through kick-out processes, where the initial defect exchanges roles with a lattice halide. So ultimately it is the structure and composition of the lattice that determines the migration, not the halide species of the initial defect.   

\subsubsection{Halide interstitials}

\begin{figure}[h!]
  \centering
    \includegraphics[]{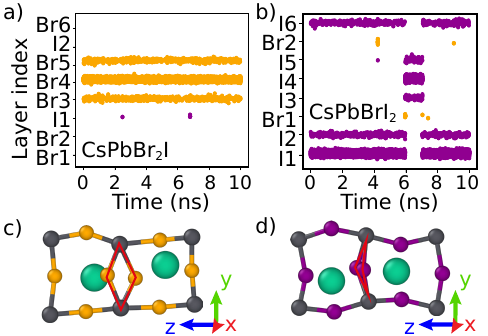}
  \caption{Position vs. simulation time of interstitial I$_\mathrm{X}$ in a) od-CsPbBr$_2$I and b) od-CsPbBrI$_2$ (for position labels, see Fig. \ref{fig:structures});  yellow, purple indicate Br, I interstitials, respectively. Snap shots of c) Pb-Br$_2$-Pb and d) Pb-Br$_2$-Pb double bridge interstitial geometries in od-CsPbBr$_2$I.}
  \label{fig:halide_int_mechanism}
\end{figure}

As a first step we focus on the anisotropy of diffusion of the halide interstitial. As shown in Fig.~\ref{fig:halide_int_mechanism}a) and b), we see that the interstitial in both od-CsPbBr$_2$I and od-CsPbBrI$_2$ is largely confined to the majority layers, i.e., Br layers and I layers respectively. As these majority layers are subject to opposite strain, i.e. tensile on the Br layer and compressive on the I layer, strain cannot be the governing factor here. 

\begin{table*}[h!]
\caption{The elementally resolved diffusion coefficients $ D_\mathrm{X}$, the time fraction of the interstitial bridge type and of the layer type population along the $x$-axis during the simulation.}
\label{tab:interstitial_values}
\centering
\small
\begin{tabular}{|l|c|c||c|c|c||c|c|c|}
\hline
\multicolumn{1}{|c|}{Structure} & \multicolumn{2}{c||}{$ D_\mathrm{X}$ ($\times 10^{-8}$ cm$^2\,$s$^{-1}$)} & \multicolumn{3}{c||}{Interstitial bridge type} & \multicolumn{3}{c|}{Layer type population} \\
\hline
 & $X$ = Br & $X$ = I  & $p_\mathrm{2I}$ (\%) & $p_\mathrm{BrI}$ (\%) & $p_\mathrm{2Br}$ (\%) & $p_\mathrm{Pb}$ (\%) & $p_\mathrm{Cs}$ (\%) & $p_\mathrm{Pb}/p_\mathrm{Cs}$ \\
\hline
$\alpha$-CsPbBr$_3$     & 24.4 & - & - & - & 100 & 67.49 & 32.51 & 2\\
$\alpha$-CsPbI$_3$      & - & 29.7 & 100 & - & - & 66.44 & 33.56 & 2\\
\hline
\hline
od-CsPbBr$_2$I      & 10.2 & 0.3 & 0 & 0.02 & 99.98 & 93.65 & 6.35 & 14.7\\
ud-CsPbBr$_2$I      & 10.7 & 0.1 & 0 & 0.12 & 99.87 & 68.13 & 31.87 & 2.1\\
\hline
od-CsPbBrI$_2$      & 0.2 & 6.0 & 99.33 & 0.67 & 0 & 92.28 & 7.72 & 12\\
ud-CsPbBrI$_2$      & 0.4 & 12.5 & 90.25 & 9.64 & 0.1 & 79.4 & 20.6 & 3.9\\
\hline
\end{tabular}
\end{table*}

We track the identity of the interstitial over time in Tab.~\ref{tab:interstitial_values}. In CsPbBr$_2$I, the interstitial is a 2Br bridge most of the time, whereas in CsPbBrI$_2$ it is an 2I bridge. 
Note that as the halide interstitial mainly migrates via a kick-out mechanism, it can easily change character over time, i.e., a Br (I) interstitial can kick out a lattice I (Br), where the latter then becomes the interstitial. We thus observe that in the interstitial adopts the character of the majority halide, i.e., Br in CsPbBr$_2$I and I in CsPbBrI$_2$.

We argue that this adoption is associated with the preferred bonding configurations of the interstitial. The most favored position of a halide interstitial is a bridge site between two neighboring Pb sites, where in fact it then forms a double bridge together with the lattice halide that is already there.\cite{Tyagi2025} The halide interstitial preferentially forms a double bridge with a halide of the same kind, i.e., Br with Br and I with I, see Tab.~\ref{tab:interstitial_values}. This observation does not depend on whether the Br and I in the CsPbBr$_2$I and CsPbBrI$_2$ materials are ordered in layers, or the materials are random alloys. It follows that the preference is dictated by the local chemical bonding.

This analysis is corroborated by an inspection of the interstitial position with respect to the metal cations over time, see Tab.~\ref{tab:interstitial_values}. One can interpret the perovskite structure as a (100) stacking PbX$_2$ and CsX layers.  
If the interstitial would have no preference while migrating, we would see a $p_\mathrm{Pb}/p_\mathrm{Cs}$ ratio of two. However, we find significantly larger values in CsPbBr$_2$I of 14.7 and in CsPbBrI$_2$ of 12.0, which indicate that the migration is dictated by the chemical bonding pattern. 

The favored bridge positions of interstitials bring about distinct diffusion paths. Fig.~\ref{fig:halide_int_mechanism}c) shows the interstitial Br in od-CsPbBr$_2$I in a characteristic Pb-Br$_2$-Pb bridge position together with a lattice Br. Hopping to neighboring site involves only a short distance and can take place unhindered in all possible directions. Migration within the Br layers then proceeds as a sequence of such nearest neighbor hopping events.

This mechanism is suppressed in the I layers within od-CsPbBr$_2$I, where the strain on the iodine layer also affects the interstitial geometry as shown in Fig.~\ref{fig:halide_int_mechanism}d). The Pb-I$_2$-Pb bridge is no longer free to rotate in all space directions equally, but has a preference for the two I to be stacked in the $x$-direction. This hinders jumps of the interstitial in the $y$-direction.

Tab.~\ref{tab:interstitial_values} lists the extracted diffusion coefficients for interstitial diffusion. 
In general, mixing halides reduces the interstitial mobility. In all structures, the overall diffusion coefficient is reduced by more than 50\%. The lowest value is observed in od-CsPbBrI$_2$, where the interstitial is mostly migrating in the Pb-I layers. Here, the favored bridge position, as shown in Fig.~\ref{fig:halide_int_mechanism}d) suppresses migration within the Pb-I layer, as discussed above.

In ordered and unordered structures of CsPbBr$_2$I and CsPbBrI$_2$, the elementally resolved diffusion coefficients are very similar. Thus in conclusion, the dominant steering feature for diffusion of halide interstitials in mixed halide perovskites is the halide composition.

\subsubsection{Halide Vacancies}
\begin{figure}[h!]
  \centering
    \includegraphics[]{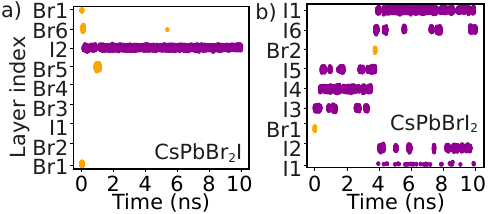}
  \caption{Position of vacancy V$_\mathrm{X}$ vs. simulation time in in a) od-CsPbBr$_2$I and b) od-CsPbBrI$_2$ (for position labels, see Fig. \ref{fig:structures}); yellow, purple indicate positions in Br and I layers, respectively.}
  \label{fig:halide_vac_X}
\end{figure}

Fig.~\ref{fig:halide_vac_X} shows the location of the vacancy in the od-CsPbBr$_2$I and od-CsPbBrI$_2$ lattices during the simulation. It is clear that the vacancy has a strong preference for migrating in the iodine layers. 
In CsPbBr$_2$I the migration is two-dimensional in a single Pb-I layer, whereas in CsPbBrI$_2$ it is somewhat more spread out but still very anisotropic. In the latter, the Pb-I layers are more populated than the Cs-I layers. 

The migration pattern is consistent with the favored position of the vacancy on the basis of its energy landscape. Tab.~\ref{tab:Br_vac_GO_energies} shows the total energies calculated for the vacancy at different positions in the CsPbBr$_2$I and CsPbBrI$_2$ lattices (with details in SI~Note~2.1). It clearly demonstrates that the energies for vacancies in Br layers are higher than in I layers. Furthermore, the vacancy prefers the Pb-I layers over Cs-I layers, which is consistent with the observation that the vacancy migrates predominantly in the Pb-I layers, and not in the Cs-I layers.

These findings are in line with observations of vacancy diffusion in simple atomic lattices, where compressive strain enhances diffusion of vacancies, and tensile strain hinders it.\cite{Tyagi2026} The commonly expressed rationale behind this statement, as noted in Ref. 42, is that compressive (tensile) strain stabilizes (destabilizes) the vacancy, and decreases (increases)its migration barrier. Our calculations (Tab.~\ref{tab:Br_vac_GO_energies} and Tab.~S6 in SI~Note~2.1) show that vacancies are energetically favored in Pb-I layers, consistent with the stabilizing effect of compressive strain in these regions. As discussed above, in layered CsPbBr$_\mathrm{x}$I$_\mathrm{3-x}$ compounds the I layers are under compressive stress, whereas the Br layers are under tensile stress.

\begin{table}
\centering
\caption{Total energies and standard error of the mean in CsPbBr$_2$I and CsPbBrI$_2$ with a halide vacancy in different positions.}
\label{tab:Br_vac_GO_energies}
\begin{tabular}{|l|r|l|r|}
\hline
\multicolumn{2}{|c|}{CsPbBr$_2$I} &  \multicolumn{2}{c|}{CsPbBrI$_2$} \\
\hline
Position & $\mathrm{\Delta}\mathrm{E}_\mathrm{tot}$ (eV) & Position & $\mathrm{\Delta}\mathrm{E}_\mathrm{tot}$ (eV)  \\
\hline
Cs-Br &   0.18 $\pm 0.03$  & Cs-I         & 0.04 $\pm 0.02$\\
Pb-Br &   0.16 $\pm 0.03$ & Pb-Br        & 0.18  $\pm 0.02$\\
Pb-I &    0 $\pm 0.03$ & Pb-I         & 0 $\pm 0.02$ \\
\hline
\end{tabular}
\end{table}

Tab.~\ref{tab:vacancy_D} shows the elementally resolved diffusion coefficients for a halide vacancy in CsPbBr$_2$I and CsPbBrI$_2$ in the ordered, as well as in the random structures, compared to $\alpha$-CsPbBr$_3$ and $\alpha$-CsPbI$_3$. The results clearly show that vacancy-mediated mass transport in mixed-halide phases is predominantly iodine migration. This holds for the ordered as well as the unordered structures. However, the difference between the diffusion coefficients of iodine and bromine migration in ordered structures is one to two orders of magnitude, whereas in unordered structures it is only a factor of 1.2-3.6.

In ordered structures, the layered arrangement combined with the preference of the vacancy for Pb-I sites strongly favors iodine migration over bromine migration. In unordered structures, asymmetric local environments enable both halides to participate in vacancy diffusion, though iodine remains dominant. This suggests that the energetic stability of vacancy sites in iodine-rich regions plays a key role in steering vacancy-mediated transport, but the precise impact of strain and chemical bonding on migration barriers remains to be quantified.

 \begin{table}[h!]
  \caption{Elementally resolved diffusion coefficients $ D_\mathrm{X}$ for a halide vacancy in CsPbBr$_\mathrm{x}$I$_\mathrm{3-x}$. In all systems a single bromine has been removed, except for the pristine $\alpha$-CsPbI$_3$, where a single iodine has been removed.}
  \label{tab:vacancy_D}
  \centering
  \small
  \begin{tabular}{|l|c|c|}
\hline
\multicolumn{1}{|c|}{} & \multicolumn{2}{c|}{$ D_\mathrm{X}$ ( $\times 10^{-9}$ cm$^2\,$s$^{-1}$)} \\
\hline
 Structure & X = Br & X = I  \\
\hline
$\alpha$-CsPbBr$_3$   & 3.9 & -   \\
$\alpha$-CsPbI$_3$    & - &  9.7 \\
\hline
\hline
od-CsPbBr$_2$I    & 0.0 & 23.8 \\
ud-CsPbBr$_2$I    &  2.3 & 8.3 \\
\hline
od-CsPbBrI$_2$    & 0.9 & 13.6 \\
ud-CsPbBrI$_2$    & 6.3 & 7.4 \\
      \hline
    \end{tabular}
  \end{table}

We observe a faster diffusion in ordered structures, compared to the pristine and unordered structures. In addition, we see a trend in the mixed structures where a higher Br content leads to an increase in iodine diffusion coefficient. This is again consistent with the strain argument mentioned above. Compressive strain promotes iodine diffusion. In unordered structures with a higher Br content the compressive strain on the I distributed in the lattice is largest among the systems we have considered here.

\section{Summary and conclusions}

In this work, we studied defect migration in mixed-halide perovskites CsPbBr$_\mathrm{x}$I$_\mathrm{3-x}$ with different halide ordering by means of molecular dynamics simulations using a reactive force field (ReaxFF). In particular, we investigated atomically layered Br and I structures similar to the ones realized experimentally.\cite{Deng2023} 

Using dynamic Voronoi analysis,\cite{Goldmann2025,Krenzer2023} we characterized the defect migration. We  found that in CsPbBr$_2$I and CsPbBrI$_2$ defect migration is very anisotropic. For Cs and halide interstitials and vacancies the rate of defect migration parallel to the layers is comparable to that observed in single-halide or random mixed-halide perovskites. However, defect migration perpendicular to ordered Br and I layers is markedly slower.

We established the factors governing the anisotropic migration in these perovskites. For Cs vacancies and interstitials the anisotropy is fully determined by the anisotropic strain introduced by ordering halides in layers in CsPbBr$_\mathrm{x}$I$_\mathrm{3-x}$. The compressive strain introduced in the latter freezes a tilting pattern in the PbX$_6$ octahedra, severely hindering Cs diffusion. At the same time, this compressive strain stabilizes the halide vacancies within the Pb-I layers, so the latter have a strong tendency to migrate within those. 

While influenced by strain, migration of halide interstitials is also governed by chemical bonding. In its most stable configuration, a halide interstitial forms a double bridge between two Pb sites with another (a lattice) halide. The system has a strong preference for a double bridge pair of the same halide, Pb-II-Pb or Pb-BrBr-Pb. Diffusion is then governed by forming such pairs along the migration path, which results in a strong anisotropy in layered perovskites. Applying a rare event sampling approach may further clarify the local initialization conditions governing defect migration.\cite{Wilke2025,Roet2021,Zhang2024}

In short, we established that strain engineering, here done through halide layering, can confine defect motion to specific crystallographic planes. In a device under operating conditions, ionic motion can be influenced by electric fields, which may have unwanted consequences such as hysteresis in current-voltage curves, or even enhanced degradation at the perovskite surfaces and interfaces. In layered perovskites such motion can be suppressed by having the layers perpendicular to the electric field. In general, the current work encourages further effort towards strain engineering in perovskite solar cells.

  \begin{acknowledgement}

S.T. and K.W. acknowledge funding from Vidi (project no. VI.Vidi.213.091) from the Dutch Research Council (NWO). This work used the Dutch national e-infrastructure with the support of the SURF Cooperative using grant no. EINF-11274. We thank Benjamin J. Morgan for helpful discussions regarding tracking mobile defects using site-projection methods. We thank Viren Tyagi for fruitful exchanges on molecular simulations of defects.
  \end{acknowledgement}
  \begin{suppinfo}
The Supporting Information contains: directional diffusion coefficients and orientational statistics for Cs vacancies and interstitials; trajectory data, bridge-configuration statistics, and directional diffusion coefficients for halide vacancies and interstitials with iodine defects; vacancy site energies from geometry optimization; time-averaged octahedral tilt patterns and their temperature dependence for layered and biaxially strained CsPbBr$_x$I$_{3-x}$ structures; finite-size convergence tests comparing $4\times4\times4$ and $6\times6\times6$ supercells; Arrhenius activation energies and pre-exponential factors for all defect types; ReaxFF force validation against DFT reference calculations; and ReaxFF vss DFT comparison of defect geometries (Pb--Pb distances, volumetric strain, and Cs--Cs distances) for Cs vacancies and interstitials.
  \end{suppinfo}
\clearpage

  \bibliography{Paper_Layers}

\end{document}


\clearpage

\tableofcontents

\noindent

\clearpage

\section{SI Note: Cs defects}
\begin{figure}[h!]
    \centering
      \includegraphics[]{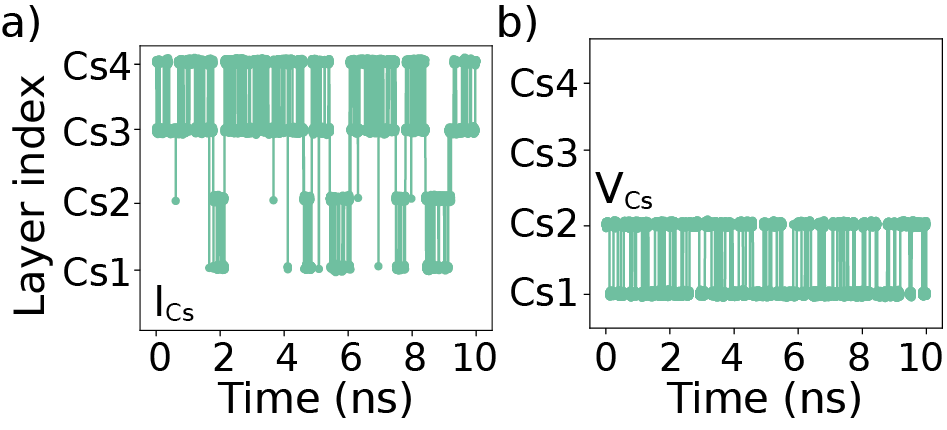} \\
    \caption{Layer index of a) the interstitial I$_\mathrm{Cs}$ and b) the vacancy V$_\mathrm{Cs}$ vs. simulation time in od-CsPbBrI$_2$. For the layer indices, see Fig. 1 in the main text.}
    \label{fig:cs_jumps_across_x}
  \end{figure}

This section provides additional data for Cs vacancy and interstitial diffusion, complementing the results presented in Section~3.1 of the main text. The Br and I layers in od-CsPbBr$_\mathrm{x}$I$_\mathrm{3-x}$ structures are parallel to the $yz$-plane. In Fig.~\ref{fig:cs_jumps_across_x} we show the jumping of Cs along the $x$-axis, in Tab.~\ref{tab:cs_dif} its the respective directional diffusion. Furthermore, we present the complementary values for directional orientation of I$_\mathrm{Cs}$ in further compounds in Tab.~\ref{tab:cs_int_orientation}. 

\begin{table*}
\caption{Elementally resolved total diffusion coefficients $D_\mathrm{X}$ and along the cartesian axes $x$, $y$, and $z$ in $\times 10^{-8}$ cm$^2\,$s$^{-1}$ for systems with a vacancy V$_\mathrm{Cs}$ or an interstitial I$_\mathrm{Cs}$ at 700~K.}
\label{tab:cs_dif}

  \centering
\begin{tabular}{|l|c|c|c|c||c|c|c|c|}
\hline
\multicolumn{1}{|c|}{} & \multicolumn{4}{c||}{V$_\mathrm{Cs}$} & \multicolumn{4}{c|}{I$_\mathrm{Cs}$} \\
\hline
$D$ in [$\times 10^{-8}$ cm$^2/$s$^{-1}$] & $D_\mathrm{tot}$ & $D_x$ & $D_y$ & $D_z$ & $D_\mathrm{tot}$ & $D_x$ & $D_y$ & $D_z$  \\
\hline
$\alpha$-CsPbBr$_3$         & 1.2 & 2.0 & 1.2 & 0.7 & 70.1 & 97.5 & 62.3 & 50.4 \\
strained CsPbBr$_3$         & 0.9 & 1.5 & 0.3 & 0.9 & 61.7 & 101.0 & 31.0 & 50.6 \\
od-CsPbBrI$_2$     & 2.2 & 1.9 & 2.3 & 3.1 & 26.1 & 13.9 & 33.9 & 30.9 \\
      \hline
    \end{tabular}
  \end{table*}

\begin{table}[h!]
\centering
\caption{The fraction of time that I$_\mathrm{Cs}$ is oriented along the $x$-axis and perpendicular to it.}
\label{tab:cs_int_orientation}
\begin{tabular}{|l|c|c|}
\hline
Config & $\perp x$ (\%) & $\parallel x$ (\%) \\
\hline
\hline
ud-CsPbBrI$_2$  & 31.2 & 68.8\\
od-CsPbBrI$_2$  & 35.5 & 64.5\\
ud-CsPbBr$_2$I  & 32.5 & 67.5\\
\hline
$\alpha$-CsPbBr$_3$  & 32.3 & 67.7\\
strained CsPbBr$_3$  & 42.6 & 57.4\\

\hline
\end{tabular}
\end{table}
\newpage

\section{SI Note: Halide defects}
\begin{figure}[h!]
  \centering
    \includegraphics[]{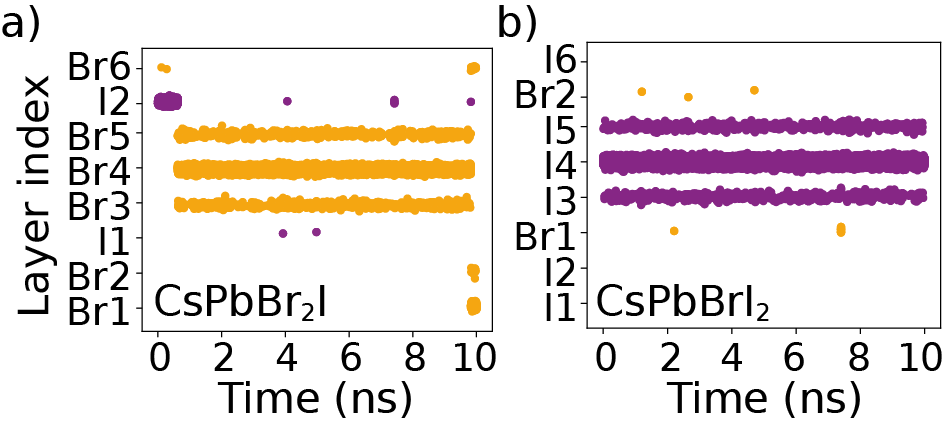}
  \caption{Layer index of I$_\mathrm{X}$ vs. simulation time in yellow when in Br layer, purple when in I layer, in a) od-CsPbBr$_2$I and b) od-CsPbBrI$_2$, starting with a single additional iodine interstitial. For the layer indices, see Fig. 1 in the main text.}
  \label{fig:halide_int_mechanism}
\end{figure}

\begin{figure}[h!]
  \centering
    \includegraphics[]{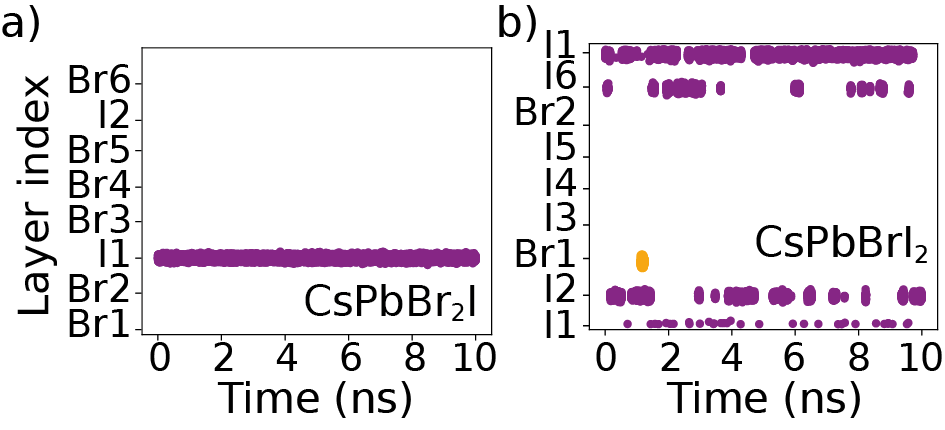}
  \caption{Layer index of V$_\mathrm{X}$ vs. simulation time in yellow when in Br layer, purple when in I layer in a) od-CsPbBr$_2$I and b) od-CsPbBrI$_2$, starting with a single iodine vacancy. For the layer indices, see Fig. 1 in the main text.}
  \label{fig:halide_vac_X}
\end{figure}

This section provides additional trajectory data and diffusion coefficients for halide vacancies and interstitials with an iodine defect, complementing the bromine-defect results presented in Section~3.2 of the main text. In Fig.~\ref{fig:halide_int_mechanism} and Fig.~\ref{fig:halide_vac_X} we present the layer index of the position of the defect, within od-CsPbBr$_2$I and od-CsPbBrI$_2$ with a single iodine added or missing. The identity of the interstitial is presented in Tab.~\ref{tab:halide_interstitial}, complementary to Tab.~2 in the main text. Directional diffusion for all compounds with halide interstitials are presented in Tab.~\ref{tab:rates_in_layers_int} and vacancies in Tab.~\ref{tab:rates_in_layers_vac}.

\begin{table}[h!]
\centering
\caption{Percentage of time an interstitial is part of a bridge configuration with two I ($p_\mathrm{2I}$), one Br and one I ($p_\mathrm{BrI}$) and two Br ($p_\mathrm{2Br}$) in CsPbBr$_\mathrm{x}$I$_\mathrm{3-x}$, starting  from a single added iodine. The interstitial type is identified as the center of a doubly occupied Voronoi cell.}
\label{tab:halide_interstitial}
\begin{tabular}{|lccc|}
\hline
Config & $p_\mathrm{2I}$ (\%) & $p_\mathrm{BrI}$ (\%) & $p_\mathrm{2Br}$ (\%) \\
\hline
\multicolumn{4}{|l|}{CsPbBrI$_2$} \\
ud- & 97.1 & 2.9 & 0 \\
od- & 99.95 & 0.5 & 0 \\
\hline
\multicolumn{4}{|l|}{CsPbBr$_2$I} \\
ud- & 0 & 0.03 & 99.97 \\
od- & 6.17 & 0.41 & 93.42 \\
\hline
\end{tabular}
\end{table}

\begin{table*}[h!]
\centering
\caption{Elementally resolved total diffusion coefficients $D_\mathrm{X}$ and along the cartesian axes $x$, $y$, and $z$ in $\times 10^{-8}$ cm$^2\,$s$^{-1}$ for CsPbBr$_\mathrm{x}$I$_\mathrm{3-x}$ structures with I$_\mathrm{X}$  at 700~K, for both an additional iodine and an additional bromine.}
\label{tab:rates_in_layers_int}
\begin{tabular}{|l l|c|c|c|c|}
\hline
\multicolumn{2}{|c}{} &  \multicolumn{4}{c|}{Interstitial} \\
\hline
\multicolumn{2}{|l|}{$D$ in [$\times 10^{-8}$ cm$^2\,$s$^{-1}$]}& $D_\mathrm{tot}$ & $D_x$ & $D_y$ & $D_z$ \\
\hline
\multicolumn{2}{|c}{} & \multicolumn{4}{c|}{Additional Iodine} \\
\hline
\multicolumn{2}{|l|}{$\alpha$-CsPbI$_3$} & 30.3 & 30.1 & 30.0 & 30.7 \\
\multicolumn{2}{|l|}{strained CsPbI$_3$} & 29.0 & 30.6 & 28.3 & 28.1 \\
\hline
od-CsPbBr$_2$I & X$=$I & 0.4 & 0.0 & 0.6 & 0.6 \\
& X$=$Br & 10.5 & 2.1 & 14.6 & 14.0 \\
\hline
od-CsPbBrI$_2$ & X$=$I & 5.2 & 2.5 & 5.7 & 7.2 \\
& X$=$Br & 0.0 & 0.0 & 0.0 & 0.0 \\
\hline
\multicolumn{2}{|c}{} & \multicolumn{4}{c|}{Additional Bromine} \\  
\hline
\multicolumn{2}{|l|}{$\alpha$-CsPbBr$_3$} & 24.4 & 22.8 & 24.6 & 25.7 \\
\multicolumn{2}{|l|}{strained CsPbBr$_3$} & 24.1 & 21.7 & 21.6 & 28.9 \\
\hline
od-CsPbBr$_2$I & X$=$Br & 10.2 & 3.1 & 18.1 & 12.0 \\
& X$=$I & 0.3 & 0.0 & 0.5 & 0.5 \\
\hline
od-CsPbBrI$_2$ & X$=$Br & 0.2 & 0.3 & 0.1 & 0.1 \\
& X$=$I & 6.0 & 3.5 & 6.7 & 7.8 \\
      \hline
    \end{tabular}
  \end{table*}

  \begin{table*}[h!]
\centering
\caption{Elementally resolved total diffusion coefficients $D_\mathrm{X}$ and along the cartesian axes $x$, $y$, and $z$ in $\times 10^{-9}$ cm$^2\,$s$^{-1}$ for CsPbBr$_\mathrm{x}$I$_\mathrm{3-x}$ structures with V$_\mathrm{X}$  at 700~K, for both a missing iodine and a missing bromine.}
\label{tab:rates_in_layers_vac}
\begin{tabular}{|l l|c|c|c|c|}
\hline
\multicolumn{2}{|c}{} & \multicolumn{4}{c|}{Vacancy} \\
\hline
\multicolumn{2}{|l|}{$D$ in [$\times 10^{-9}$ cm$^2\,$s$^{-1}$]} & $D_\mathrm{tot}$ & $D_x$ & $D_y$ & $D_z$\\
\hline
\multicolumn{2}{|c}{} & \multicolumn{4}{c|}{Missing Iodine} \\
\hline
\multicolumn{2}{|l|}{$\alpha$-CsPbI$_3$         } & 9.7 & 8.9 & 9.2 & 11.0 \\
\multicolumn{2}{|l|}{strained CsPbI$_3$     } & 13.0 & 11.5 & 14.4 & 12.9 \\
\hline
od-CsPbBr$_2$I & X$=$I & 20.0 & 0.0 & 35.5 & 24.1 \\
& X$=$Br           & 0.0 & 0.0 & 0.1 & 0.1 \\
\hline
od-CsPbBrI$_2$   & X$=$I & 14.1 & 13.9 & 12.3 & 16.1 \\
& X$=$Br                 & 0.0 & 0.0 & 0.0 & 0.0 \\
\hline
\multicolumn{2}{|c}{} & \multicolumn{4}{c|}{Missing Bromine} \\  
\hline
\multicolumn{2}{|l|}{$\alpha$-CsPbBr$_3$} & 3.9 & 5.0 & 3.3 & 3.4 \\
\multicolumn{2}{|l|}{strained CsPbBr$_3$    } & 5.2 & 2.4 & 6.6 & 6.5 \\
\hline
od-CsPbBr$_2$I & X$=$Br & 0.1 & 0.1 & 0.1 & 0.1 \\
& X$=$I                  & 23.8 & 0.2 & 31.7 & 40.2 \\
\hline
od-CsPbBrI$_2$ & X$=$Br & 0.9 & 1.3 & 0.0 & 1.3 \\
& X$=$I                  & 13.6 & 15.9 & 12.1 & 12.9 \\
      \hline
    \end{tabular}
  \end{table*}

\clearpage
\subsection{Halide vacancy energy landscapes}
In the following, we provide details on the procedure used to obtain site-dependent energies for halide vacancies in CsPbBr$_\mathrm{x}$I$_\mathrm{3-x}$, complementing the results presented in Tab.~3 in Section~3.2.2 of the main text. Site-dependent energies for halide vacancies were obtained by geometry optimization of all distinct vacancy positions in the 4$\times$4$\times$4 supercell, using the same software and force field as the MD simulations. Each halide site in the pristine supercell was probed for the vacancy,
yielding three symmetry-inequivalent positions per structure (Pb-Br, Pb-I, and Cs-Br/Cs-I), each with 64 equivalent instances. Prior to optimization, a short $NVT$-MD run of 1.25~ps at 500~K with a Berendsen thermostat ($\tau_T = 100$~fs) was performed to allow for local relaxation around the vacancy. Geometry optimizations were then carried out using a conjugate gradient algorithm (``Good'' quality setting in AMS) at fixed cell volume. Reported energies are averages over all equivalent positions; error bars represent the standard error of the mean. In cases where the vacancy migrated to a neighboring layer during optimization, the displaced species was constrained to its initial neighboring-layer position. Results for a missing bromine are given in the main text; results for a missing iodine are given in Tab.~\ref{tab:I_vac_GO_energies}.

\begin{table}
\centering
\caption{Relative total energies $\Delta E_\mathrm{tot}$ of vacancy sites in CsPbBr$_2$I ($\pm 0.03$~eV) and CsPbBrI$_2$ ($\pm 0.02$~eV) for a missing iodine. Energies are referenced to the lowest-energy Pb-I site.}
\label{tab:I_vac_GO_energies}
\begin{tabular}{|l|c||l|c|}
\hline
\multicolumn{1}{|c}{Missing I in:} &  \multicolumn{1}{c||}{CsPbBr$_2$I} & \multicolumn{2}{c|}{CsPbBrI$_2$}   \\
\hline
Position & Total Energy (eV) & Position & Total Energy (eV) \\
\hline
Cs-Br &   0.14       & Cs-I         & 0.03  \\
Pb-Br &   0.14       & Pb-Br        & 0.1  \\
Pb-I &    0       & Pb-I         & 0  \\
\hline
\end{tabular}
\end{table}

\clearpage
\section{SI Note: Influence of strain}\label{sec:tilts}
\begin{figure*}[h!]
    \centering
    \includegraphics[]{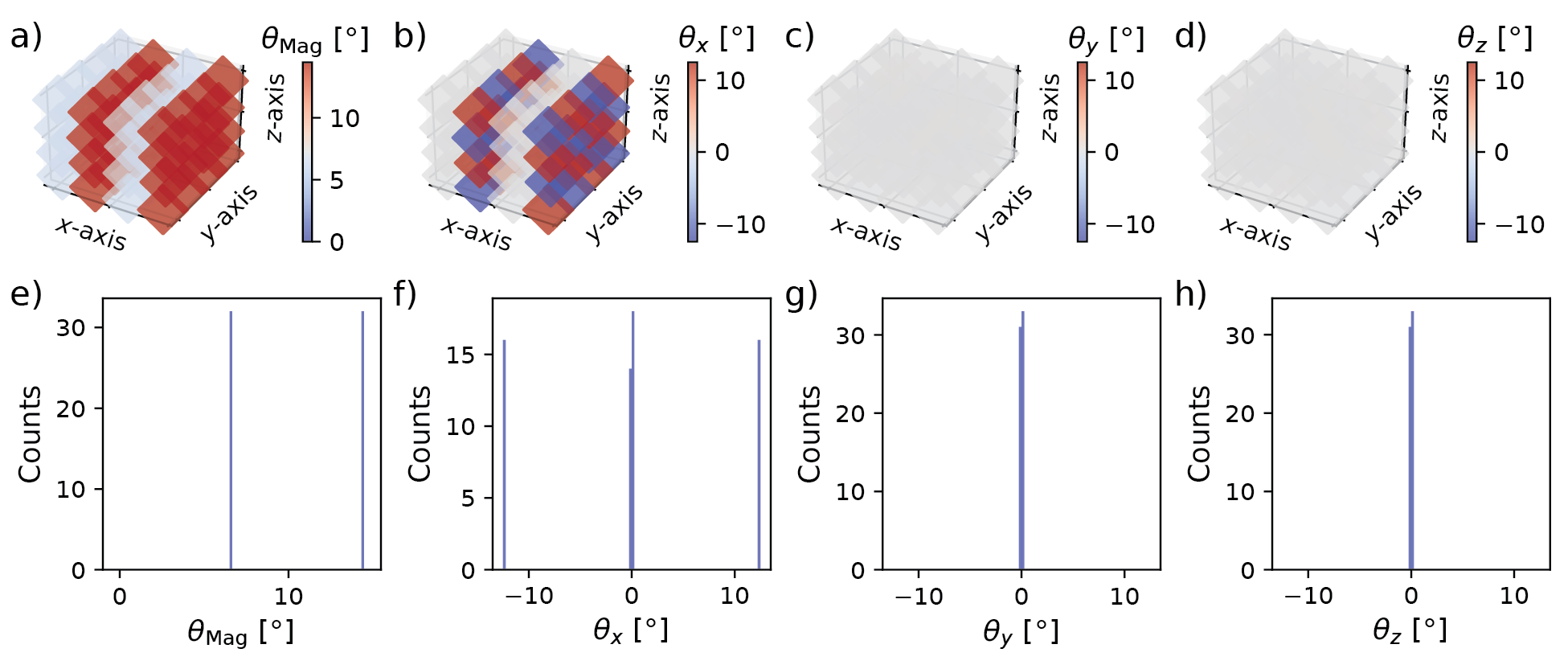}
\caption{a) Time-averaged rotation magnitude  of the PbX$_6$ octahedral tilts and b)-d) time-averaged tilt angles with respect to the cartesian axes ($\theta_x,\theta_y,\theta_z$) in layered CsPbBr$_2$I perovskites. The color scale indicates the angle (in degrees) for each octahedron. Panel e)-h) are the histograms, referring to the respective pattern above, with 100 bins. The maxima are indicating different types of octahedrons. For Br octahedrons we observe $\theta_\mathrm{Mag}\approx 6.6^{\circ}$, while $\theta_x,\theta_y,\theta_z\approx0^{\circ}$, whereas for I octahedrons $\theta_\mathrm{Mag}\approx 14.4^{\circ}$, while $\theta_y,\theta_z\approx0$ and $\theta_x\approx \pm 12.4^{\circ}$.}
    \label{fig:3-1_tilt_pattern}
  \end{figure*}

\begin{figure*}[h!]
    \centering
    \includegraphics[]{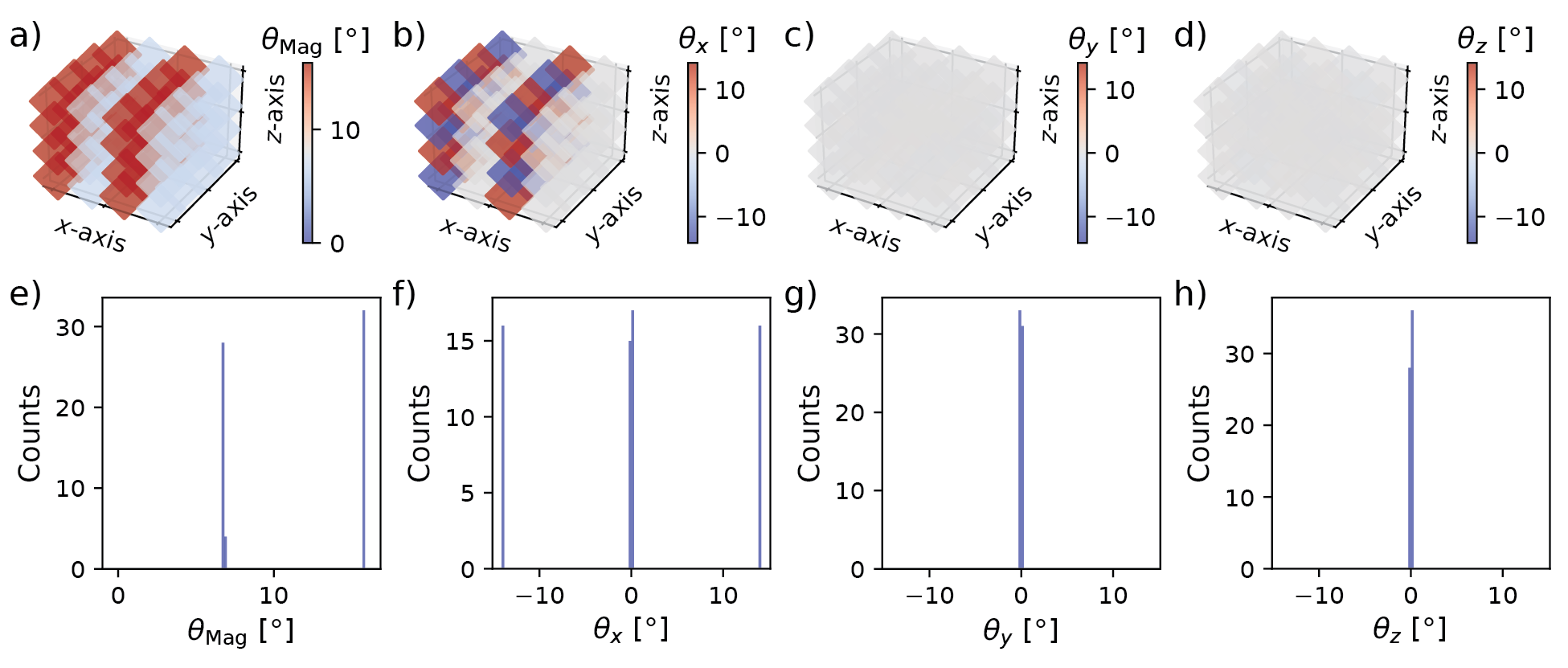}
\caption{a) Time-averaged rotation magnitude of the PbX$_6$ octahedral tilts and b)-d) time-averaged tilt angles with respect to the cartesian axes ($\theta_x,\theta_y,\theta_z$) in layered CsPbBrI$_2$  perovskites. The color scale indicates the angle (in degrees) for each octahedron. Panel e)-h) are the histograms, referring to the respective pattern above, with 100 bins. The maxima are indicating different types of octahedrons. For Br octahedrons we observe $\theta_\mathrm{Mag}\approx 6.8^{\circ}$, while $\theta_x,\theta_y,\theta_z\approx0^{\circ}$, whereas for I octahedrons $\theta_\mathrm{Mag}\approx 15.8^{\circ}$, while $\theta_y,\theta_z\approx0^{\circ}$ and $\theta_x\approx \pm 14.0^{\circ}$}
    \label{fig:1-3_tilt_pattern}
  \end{figure*}

\begin{figure}[h!]
  \centering
  \includegraphics[]{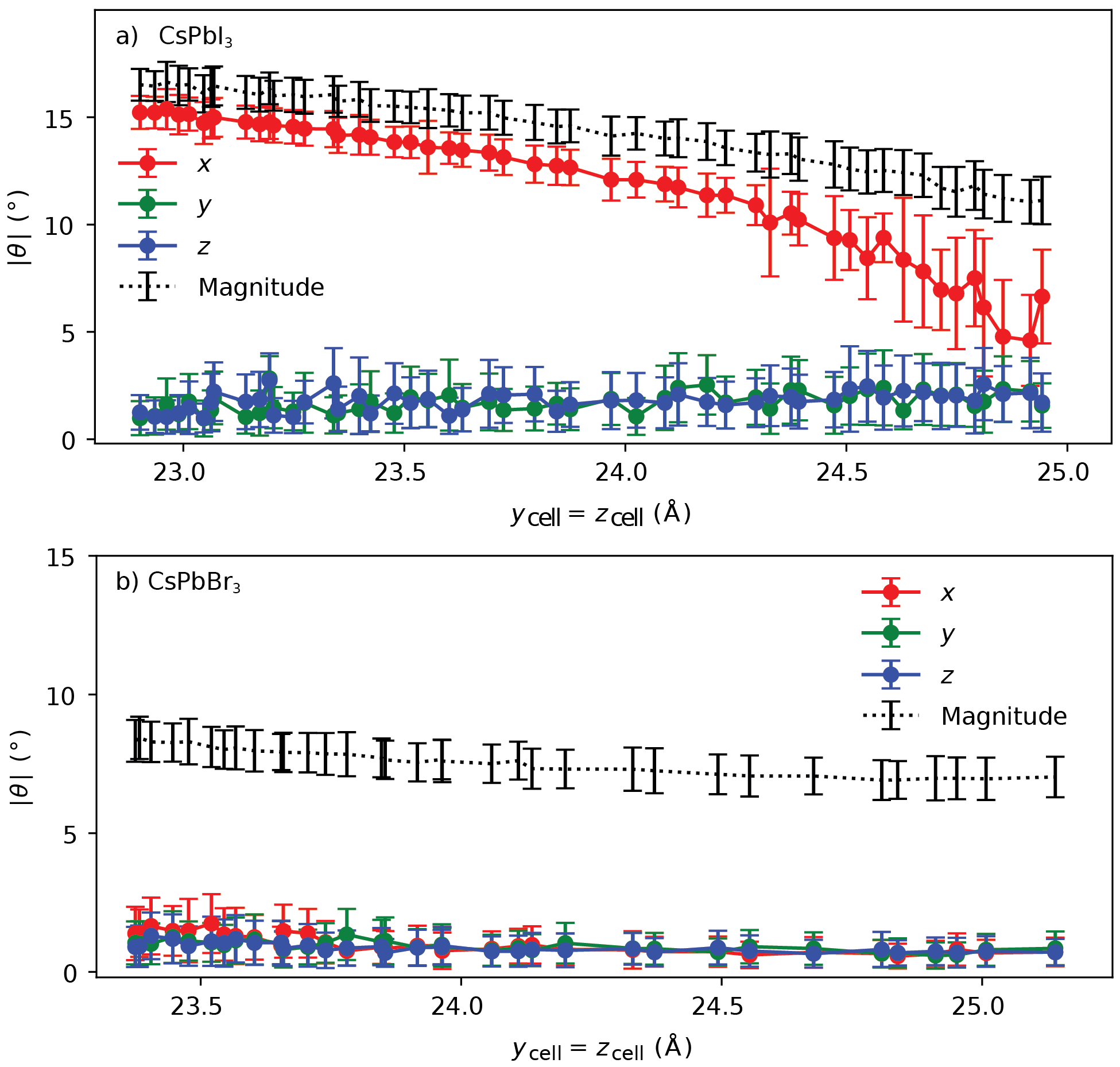}
\caption{The supercell vectors $y_\mathrm{cell} = z_\mathrm{cell}$ vs. the absolute tilt $|\theta|$ for a) CsPbI$_3$ under compressive stress and $x_\mathrm{cell}=25.53$~\AA{}; b) CsPbBr$_3$ under tensile stress and $x_\mathrm{cell}=23.83$~\AA{}. The target volumes for achieving the desired tilt patterns are highlighted.}
\label{fig:pressure_curves}
\end{figure}

\begin{figure*}[h!]
  \centering
  \includegraphics[]{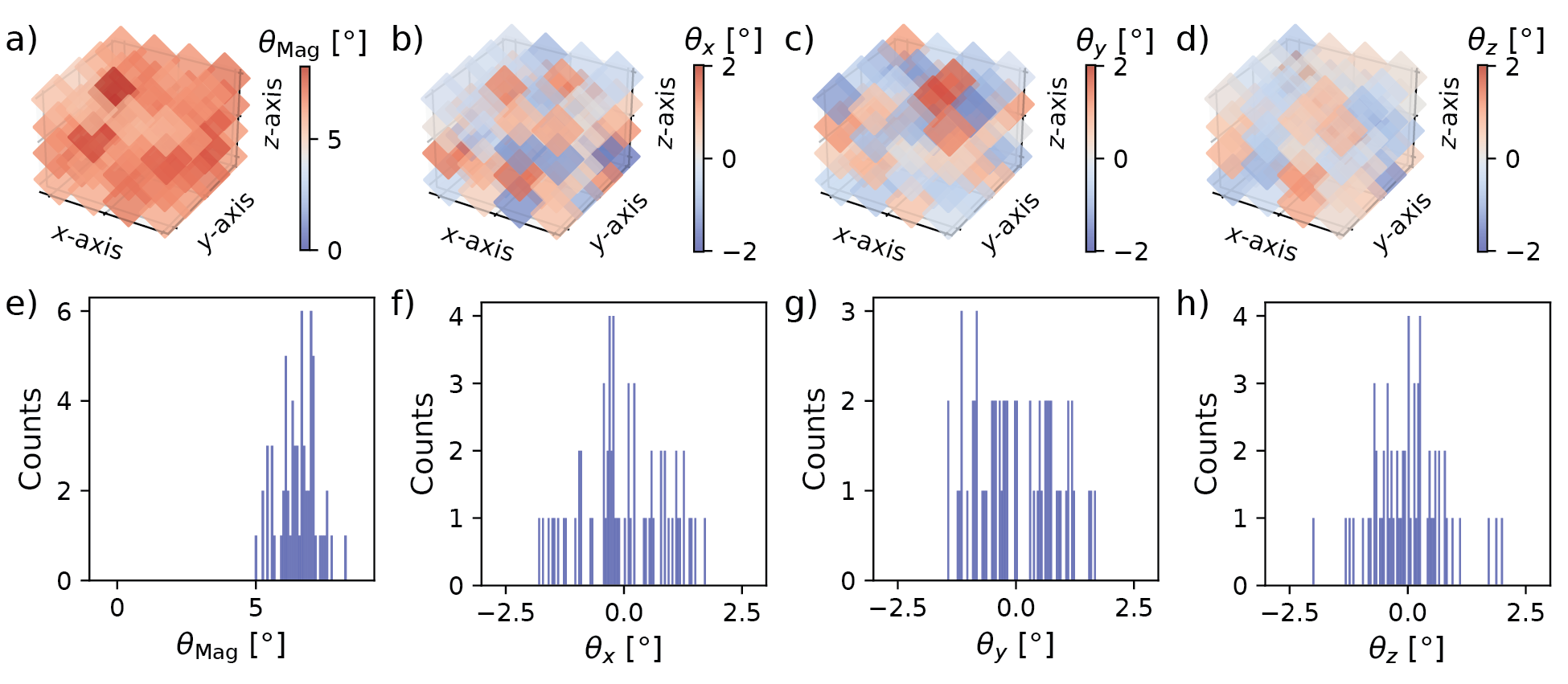}
  \caption{a) Time-averaged rotation magnitude of the PbX$_6$ octahedral tilts and b)-d) their components along the axes in CsPbBr$_3$ under biaxial compression to achieve similar tilt magnitudes as in layered CsPbBr$_2$I$_{1}$. The color scale indicates the angle (in degrees) for each octahedron. Panel e)-h) are the histograms, referring to the respective pattern above, with 100 bins. We observe $\theta_\mathrm{Mag}\approx 6.7^{\circ}$, while $\theta_x,\theta_y,\theta_z\approx0$.}
  \label{fig:strained_cspbbr3_tilt_pattern}
\end{figure*}

This section provides the full octahedral tilt pattern analysis for the layered and biaxially strained structures discussed in Section~3.1 of the main text, along with the procedure used to determine the target strain values. The rotation magnitude  $\theta_\mathrm{Mag}$ of the PbX$_6$ octahedra is octahedra is derived from the complex conjugate eigenvalues of the rotation matrix, specifically $e^{i\theta_\mathrm{Mag}}$ and $e^{-i\theta_\mathrm{Mag}}$. A detailed analysis on the octahedral tilts can be found for CsPbBr$_2$I in Fig.~\ref{fig:3-1_tilt_pattern} and for CsPbBrI$_2$ in Fig.~\ref{fig:1-3_tilt_pattern}. There was strain applied to the pristine structures of $\alpha$-CsPbI$_3$  and $\alpha$-CsPbBr$_3$ in Fig.~\ref{fig:pressure_curves} to introduce octahedral tilting. The detailed analysis of those structures can be found for CsPbI$_3$ in Fig.~\ref{fig:strained_cspbi3_tilt_pattern} and for CsPbBr$_3$ in Fig.~\ref{fig:strained_cspbbr3_tilt_pattern}. 
\subsection{Layered Structures}

Tilt directionality also shows a well-defined spatial pattern. Within the 2D Pb-X layers (parallel to the $yz$ plane), we find a checkerboard-like alternating tilt pattern of $\theta_x$ in the Pb-I layers: individual octahedra are tilted, such that every octahedron is surrounded in all in-plane directions by neighbors tilted in the opposite sense. The tilt around the $y$- and $z$-axes as well as the magnitude of the octahedral tilts and all angular components in the Pb-Br layers are close to zero, indicating a clear confinement of the distortion to the strained layers.

\subsection{Temperature Dependence}

\begin{figure*}[h!]
  
  \centering

    \includegraphics[]{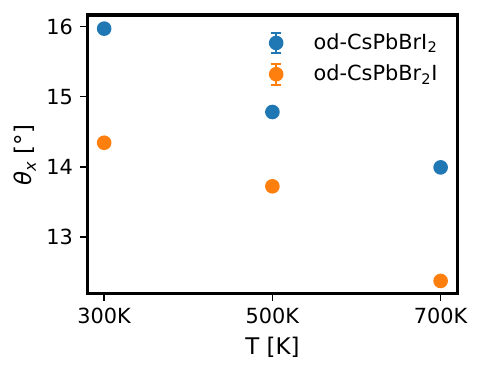}
\caption{Average absolute tilt angle $\theta_x$ of od-CsPbBr$_2$I and od-CsPbBrI$_2$ of all PbX$_2$I$_4$ octahedrons versus temperature obtained from a 2~ns simulations.}
    \label{fig:tilt_vs_T}
  \end{figure*}

In agreement with the results presented by \citeauthor{Pols2024}, we find that the octahedral tilts decrease with temperature.\cite{Pols2024} We have determined $\theta_x$ of the iodine layers in CsPbBr$_2$I and CsPbBrI$_2$ in Fig.~\ref{fig:tilt_vs_T}. The tilts have been determined from 2 ns runs of pristine systems with the same MD settings as described in the main text.

\subsection{Applied biaxial strain to pristine $\alpha$-CsPbBr$_3$ and $\alpha$-CsPbI$_3$}
To assess whether strain alone can reproduce the characteristic octahedral tilts observed in the layered structures, biaxial compression was applied in the $yz$-plane to chemically homogeneous $\alpha$-CsPbI$_3$, whereas biaxial tension was applied to $\alpha$-CsPbBr$_3$. The out-of-plane ($x$) lattice parameter was held fixed at 23.83~\AA{} for CsPbBr$_3$ and 25.53~\AA{} for CsPbI$_3$ throughout. Compressive (tensile) stress along the $y$- and $z$-directions was applied using the same ReaxFF force field of Ref.~\citenum{Pols2024}.

For each stress value, the following procedure was applied. A fresh supercell was pre-relaxed with a short $NVT$-MD run of 1.25~ps at 700~K. The cell was then compressed (CsPbI$_3$) or expanded (CsPbBr$_3$) equally in the $y$- and $z$-directions via a geometry optimization with only the $x$-dimension fixed, using a stress tensor of the form $[0, p, p, 0, 0, 0]$. Stress was varied over 50 steps from 1 to 30000~atm for CsPbI$_3$ and from $-1$ to $-30000$~atm for CsPbBr$_3$. The resulting cell dimensions were then used as input for a 7.5~ps $NVT$-MD run (output frequency 0.5~ps), of which the final 5~ps were used for the tilt analysis. To account for fluctuations in volume and tilt, the entire procedure was repeated three times per stress value and results were averaged. The target geometry for the production MD runs was selected as the lowest-strain configuration that reproduced the approximate $\theta_x$:$\theta_y$:$\theta_z$ ratio of the layered od-structures while maintaining structural stability.

For the MD simulations in strained systems, we fixed the dimensions of CsPbBr$_3$: 23.83x24.2x24.2~$\mathrm{\AA}$, and CsPbI$_3$: 25.53x24.52x24.52~$\mathrm{\AA}$. This achieved moderate strain in the $yz$ plane to achieve the desired tilt patterns, while still maintaining the stability of the structures in longer MD runs. For enhanced visibility of the tilt patterns, we used the most strongly compressed volume for CsPbI$_3$ and the strongest stable inflation of CsPbBr$_3$ for the following figures. 

  \begin{figure*}[h!]
    \centering
    \includegraphics[]{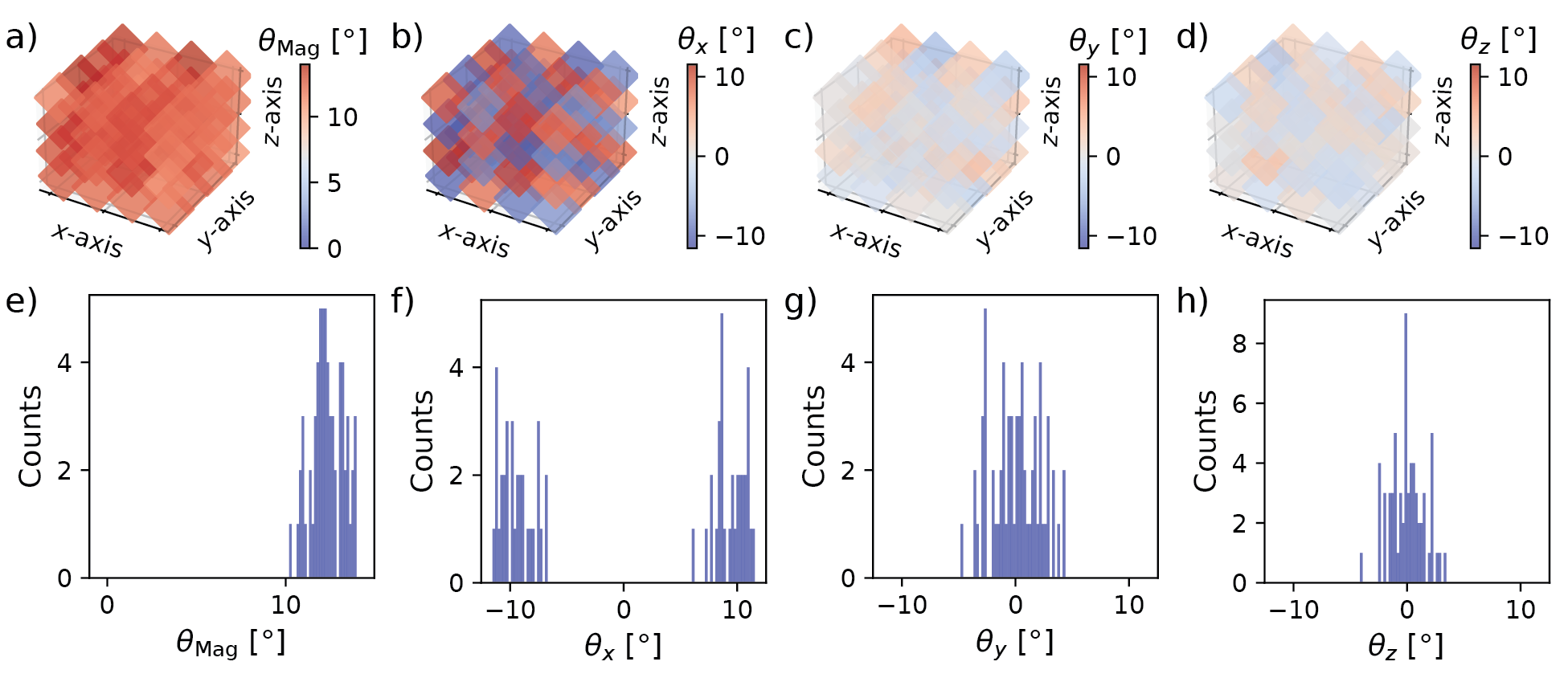}
  \caption{a) Time-averaged rotation magnitude of the PbX$_6$ octahedral tilts and b)-d) their components along the axes in CsPbI$_3$ under biaxial compression to achieve similar tilt magnitudes as in layered CsPbBr$_2$I$_{1}$. The color scale indicates the angle (in degrees) for each octahedron. Panel e)-h) are the histograms, referring to the respective pattern above, with 100 bins. We observe $\theta_\mathrm{Mag}\approx 12.6^{\circ}$, while $\theta_y,\theta_z\approx0$ and $\theta_x\approx \pm 9.3^{\circ}$}
    \label{fig:strained_cspbi3_tilt_pattern}
    \end{figure*}

As expected, the strained CsPbI$_3$-structure in Fig.~\ref{fig:strained_cspbi3_tilt_pattern} lacks the layered modulation of tilt magnitude seen in mixed-halide systems, since it is fully homogeneous and does not contain Br-rich regions. Thus, while the checkerboard pattern emerges, it is no longer separated by undistorted layers. Additionally, the tilt components perpendicular to the layers ($x$-direction), while still smaller than the in-plane components, are non-zero, in contrast to the strictly planar confinement seen in the layered architecture.
\clearpage

\section{SI Note: Validation and Benchmark of ReaxFF}

The ReaxFF model used in this work is based on the mixed-halide parameterization introduced by \citeauthor{Pols2024}.\cite{Pols2024} They provide a broad validation of the mixed-halide ReaxFF model. In brief, the force field was benchmarked against DFT for (i) atomic charges across precursor and perovskite environments, (ii) equations of state and relative phase energetics of CsPbI$_3$ and CsPbBr$_3$, (iii) mixing enthalpies of CsPbBr$_\mathrm{x}$I$_\mathrm{3-x}$, and (iv) DFT-NEB defect migration barriers (including halide vacancies and interstitials), with reported barrier agreement within 0.065~eV. The associated training-set files provided with the SI contain 544 energy targets and 89 charge targets, spanning salts and precursors (CsI/CsBr, PbI$_2$/PbBr$_2$), polymorphs of CsPbI$_3$ and CsPbBr$_3$, mixed-halide CsPbBr$_\mathrm{x}$I$_\mathrm{3-x}$ orderings, and pathways/transition-states (including $\gamma \rightarrow \delta$ conversion states). Importantly, the mixed-halide entries explicitly cover differently substituted PbX$_6$ octahedra, which are central to reproducing the octahedral-tilt behavior. This is directly relevant to the scope of Ref.~\citenum{Pols2024}, where the temperature dependence of tilt angles is evaluated as a function of Br substitution. This previously established validation provides the baseline for the present study.\\

Building on this baseline, we provide additional checks tailored to the present defect-diffusion applications in layered and strained systems. These include computational-cost and practical-accuracy considerations relative to recent NNP workflows, finite-size sensitivity tests, and activation energy. Furthermore, we compare predicted forces, tilts and defect geometries to DFT reference calculations. Due to the inclusion of halide defect validation in the original ReaxFF publication, we focus here on the Cs defect geometries, which is not covered in Ref.~\citenum{Pols2024}. The results of these additional checks are presented in the following subsections.

\subsection{Computational Cost and Accuracy Considerations}
ReaxFF provides high accuracy with acceptable computational cost compared to classical force fields while running very efficiently on CPUs, which have become increasingly accessible and affordable despite the recent GPU boom. To benchmark performance, we tested a CPU version of an Allegro/NequIP potential for negatively charged Br interstitials, published by \citeauthor{Tyagi2026}\cite{Tyagi2026} on the same hardware as ReaxFF: 16 CPUs on dual AMD Rome 7H12 processors (2.6~GHz, 280~W per socket). On a system with 321 atoms in a 4x4x4 CsPbI$_3$ cell with one bromine interstitial and nVE dynamics using ReaxFF with AMS2024\cite{Ruger_2024}, we achieved approximately 271 steps/s ($4 \times 10^{7}$ steps / $2.3 \times 10^{6}$~s), whereas the Allegro (v0.8.1)/NequIP (v0.16.1) implementation with ASE 3.27.0 achieved only approximately 2.6 steps/s ($ 10^{4}$ steps / $3.9 \times 10^{3}$~s), demonstrating a speedup of over 100-fold for ReaxFF on this CPU architecture. We note, however, that substantially higher throughput for NNPs is typically achievable on modern GPUs. In practice, such GPU workflows are more expensive and access to suitable accelerators is often limited, which motivated our CPU-based approach here. From an accuracy perspective, NNPs can provide more accurate local defect geometries because they can explicitly treat the most probable (charged) defect states, whereas in the present ReaxFF setup only neutral defects are accessible. In this work, we impose high strain levels and picked the ReaxFF approach, as we expect under these conditions strain-driven structural distortions to dominate, and geometry deviations from the neutral-defect approximation are expected to be less influential for the trends discussed. In the following subsections, we therefore provide additional validation of the ReaxFF setup in terms of size convergence, force quality, defect geometries, and migration energetics.

\subsection{Finite Size effects}

The defect-dynamics simulations in this work were performed in a $4\times4\times4$ supercell. While this cell size is sufficient to capture the qualitative migration trends discussed in the main text, it is still comparatively small and may therefore introduce finite-size effects, particularly for defect configurations with long-range elastic or electrostatic interactions. To assess the sensitivity of the defect migration behavior to supercell size, we compared the migration of Cs and halide interstitials and vacancies in CsPbI$_3$ and od-CsPbBr$_2$I to corresponding calculations in a $6\times6\times6$ supercell at 700~K. It is to note that for the comparison, we now take the concentration-independent diffusion coefficient defined as:
\begin{equation}\label{eq:arrhenius}
        D = \frac{1}{6}\lim_{t\rightarrow\infty}\frac{d}{dt}
        \Bigl\langle \sum_{i=1}^{N} 
        \left\Vert \mathbf{r}_i(t+t_0)-\mathbf{r}_i(t_0)\right\Vert^{2}
        \Bigr\rangle_{t_{0}};
         t > t_0
\end{equation}
where $N$ is the number of atoms of a mobile species in the system. This removes the dependency on the number of mobile atoms in the system from equation~(1) and (2) in the main text.

\begin{table}[h!]
\centering
\caption{Finite-size comparison of defect migration in $4\times4\times4$ and $6\times6\times6$ supercells at 700~K. Placeholder table for reporting migration values, barriers, or other relevant metrics.}
\label{tab:finite_size_effects}
\begin{tabular}{|l|l|c|c|c|c||c|c|c|c|}
\hline
\multicolumn{2}{|c|}{$D$ in [$\times 10^{-6}$ cm$^2/$s$^{-1}$] } & \multicolumn{4}{c||}{4x4x4} & \multicolumn{4}{c|}{6x6x6} \\
\hline
Structure & Defect & $D_\mathrm{tot}$ & $D_x$ & $D_y$ & $D_z$ & $D_\mathrm{tot}$ & $D_x$ & $D_y$ & $D_z$  \\
\hline
$\alpha$-CsPbI$_3$ & V$_\mathrm{X}$ & 1.85 & 1.7 & 1.75 & 2.1 & 2.61 & 2.18 & 3.03 & 2.63 \\
$\alpha$-CsPbI$_3$ & I$_\mathrm{X}$ & 58.5 & 58 & 57.9 & 59.2 & 55.6 & 53.9 & 57.5 & 55.2 \\
$\alpha$-CsPbI$_3$ & V$_\mathrm{Cs}$ & 2.86 & 2.46 & 2.28 & 3.7 & 3.45 & 2.58 & 4.84 & 2.93 \\
$\alpha$-CsPbI$_3$ & I$_\mathrm{Cs}$ & 30.9 & 26.7 & 43.4 & 22.7 & 17.8 & 23.8 & 18.1 & 12.2 \\
\hline
od-CsPbBr$_2$I & V$_\mathrm{X}$ & 1.53 & 0.02 & 2.04 & 2.58 & 1.7 & 0.00 & 2.64 & 2.45 \\
od-CsPbBr$_2$I & I$_\mathrm{X}$ & 13.4 & 4.01 & 23.6 & 15.8 & 20.4 & 6.31 & 31.3 & 23.6 \\
od-CsPbBr$_2$I & V$_\mathrm{Cs}$ & 0.50 & 0.09 & 1.03 & 0.38 & 0.44 & 0.40 & 0.57 & 0.35 \\
od-CsPbBr$_2$I & I$_\mathrm{Cs}$ & 2.64 & 0.00 & 6.57 & 1.32 & 3.46 & 0.11 & 5.48 & 4.78 \\
\hline
\end{tabular}
\end{table}

From Table~\ref{tab:finite_size_effects}, we observe that in the $4\times4\times4$ supercell the diffusion coefficients $D$ for interstitial defects (I$_\mathrm{X}$ and I$_\mathrm{Cs}$) are slightly overestimated, while those for vacancy defects are slightly underestimated, compared to the $6\times6\times6$ results. For the analysis in this work, the key importance is that in pure CsPbI$_3$ the defects explore all three spatial directions ($x$, $y$, and $z$), i.e., the migration is three-dimensional. This can be achieved equally well in the $4\times4\times4$ supercell. The suppression of migration in the $x$-direction for the od-CsPbBr$_2$I structure is also observed in both supercell sizes and the diffusion coefficients are very close. Thus, we conclude that a $4\times4\times4$ supercell is sufficient to capture the effects, we discussed in the main text. 

\subsection{Activation Energies}

For all defects, discussed in the main text, we have performed an additional, identical MD run at 500~K, to obtain an estimate of the activation energy $E_\mathrm{a}$ for diffusion. The activation energy is obtained from the Arrhenius relation:
\begin{equation}
D = D_0 \exp\left(-\frac{E_\mathrm{a}}{k_\mathrm{B}T}\right)
\end{equation}
where $D$ is the diffusion coefficient, $D_0$ is the pre-exponential factor, $k_\mathrm{B}$ is the Boltzmann constant, and $T$ is the temperature. By plotting $\ln(D)$ against $1/T$, we can extract the activation energy from the slope of the resulting line. The calculated activation energies for each defect type are summarized in Tab.~\ref{tab:activation_energies_int} and Tab.~\ref{tab:activation_energies_vac}. 

It is important to note that the diffusion coefficients reported here refer to a single run of 10~ns, and some of the defects do not have fully converged in their diffusion behavior within this time frame. Therefore, the reported activation energies should be interpreted as indicative values rather than precise measurements. The general trends are expected to be reliable, but the absolute values may vary with longer simulation times or additional runs. All of the activation energies for halide defects are in a realistic range, as summed up by \citeauthor{Arber2025a}\cite{Arber2025a} in comparison to previous experiments and calculations. Notably, in almost all od-structures, the activation energy of the dominant halide species at 700~K is significantly lower than that of the minority species, showing that the anisotropic confinement of defect diffusion is even larger at operating conditions. The exception of V$_\mathrm{X}$ in od-CsPbBr$_2$I is likely due to the negligible contribution of bromine diffusion at both temperatures, making it prone to larger errors. Although not directly obvious from the activation energies, we expect a similarly increased anisotropy for Cs defects in the od-structures, as the gate-blocking octahedral tilts are increasing, as shown in Fig.~\ref{fig:tilt_vs_T}.

In direct comparison to the neural-network potential (NNP) results of \citeauthor{Tyagi2026},\cite{Tyagi2026} we find that the activation energies for neutral halide interstitials are slightly lower in the present ReaxFF data (0.09-0.11~eV vs 0.17-0.20~eV), whereas the corresponding vacancy barriers are higher (0.51~eV vs 0.26~eV), as summarized in Tab.~\ref{tab:activation_energies_viren}. The concentration-independent pre-exponential factors, following equation~\ref{eq:arrhenius}, lie within the same order of magnitude for all cases.
\begin{table*}[h!]
\centering
\caption{Activation energies and pre-exponential factors for interstitials in CsPbBr$_\mathrm{x}$I$_\mathrm{3-x}$ perovskites. Diffusion coefficients $D$ are evaluated at 500~K and 700~K on a single MD run of 10~ns.}
\label{tab:activation_energies_int}
\begin{tabular}{|l l|c|c|c|c|}
  
\hline
\multicolumn{2}{|l|}{$D$ in [cm$^2\,$s$^{-1}$]}& $D_\mathrm{500K}$ $[\times 10^{-9}]$ & $D_\mathrm{700K}$ $[\times 10^{-9}]$ & $E_a$ [eV]  & $D_0$ $[\times 10^{-6}]$ \\
\hline
\multicolumn{2}{|c|}{} &  \multicolumn{4}{c|}{I$_\mathrm{X}$} \\
\hline
\multicolumn{2}{|c|}{} & \multicolumn{4}{c|}{Additional Iodine} \\
\hline
\multicolumn{2}{|l|}{$\alpha$-CsPbI$_3$} & 149 & 303 & 0.11 & 1.79 \\
\multicolumn{2}{|l|}{strained CsPbI$_3$} & 135 & 290 & 0.12 & 1.96 \\
\hline
\multicolumn{2}{|c|}{} & \multicolumn{4}{c|}{Additional Bromine} \\  
\hline
\multicolumn{2}{|l|}{$\alpha$-CsPbBr$_3$} & 135 & 244 & 0.09 & 1.07 \\
\multicolumn{2}{|l|}{strained CsPbBr$_3$} & 95.9 & 241 & 0.14 & 2.41 \\
\hline
od-CsPbBr$_2$I & X$=$Br & 10.8 & 102 & 0.34 & 28 \\
& X$=$I & 0.01 & 3.1 & 0.96 & 2.68$\times10^{4}$ \\
\hline
od-CsPbBrI$_2$ & X$=$Br & 0.02 & 2.01 & 0.70 & 217 \\
& X$=$I & 13.1 & 59.8 & 0.23 & 2.66 \\
\hline
ud-CsPbBr$_2$I & X$=$Br & 24.1 & 107 & 0.22 & 4.44 \\
& X$=$I & 0.00 & 1.23 & 0.90 & 3.83$\times10^{3}$ \\
\hline
ud-CsPbBrI$_2$ & X$=$Br & 0.00 & 3.62 & 1.37 & 2.56$\times10^{7}$ \\
& X$=$I & 24.7 & 125 & 0.24 & 7.2 \\
\hline
\hline
\multicolumn{2}{|c|}{} & \multicolumn{4}{c|}{I$_\mathrm{Cs}$} \\  
\hline
\multicolumn{2}{|l|}{$\alpha$-CsPbI$_3$} & 63.6 & 476 & 0.30 & 72.9 \\
\multicolumn{2}{|l|}{strained CsPbI$_3$} & 1.82 & 43.4 & 0.48 & 121 \\
\hline
\multicolumn{2}{|l|}{$\alpha$-CsPbBr$_3$} & 728 & 701 & -0.01 & 0.64 \\
\multicolumn{2}{|l|}{strained CsPbBr$_3$} & 736 & 617 & -0.03 & 0.40 \\
\hline
\multicolumn{2}{|l|}{od-CsPbBr$_2$I}  & 1.13 & 40.5 & 0.54 & 311 \\
\multicolumn{2}{|l|}{od-CsPbBrI$_2$}  & 41.5 & 261 & 0.28 & 25.9 \\
\multicolumn{2}{|l|}{ud-CsPbBr$_2$I}  & 0.96 & 138 & 0.75 & 3.45$\times10^{4}$ \\
\multicolumn{2}{|l|}{ud-CsPbBrI$_2$}  & 3.08 & 194 & 0.62 & 6.11$\times10^{3}$ \\
\hline

      \hline
    \end{tabular}
  \end{table*}

 \begin{table*}[h!]
\centering
\caption{Activation energies and pre-exponential factors for vacancies in CsPbBr$_\mathrm{x}$I$_\mathrm{3-x}$ perovskites. Diffusion coefficients $D$ are evaluated at 500~K and 700~K on a single MD run of 10~ns.}
\label{tab:activation_energies_vac}
\begin{tabular}{|l l|c|c|c|c|}
  
\hline
\multicolumn{2}{|l|}{$D$ in [cm$^2\,$s$^{-1}$]}& $D_\mathrm{500K}$ $[\times 10^{-10}]$ & $D_\mathrm{700K}$ $[\times 10^{-10}]$ & $E_a$ [eV]  & $D_0$ $[\times 10^{-6}]$ \\
\hline
\multicolumn{2}{|c|}{} &  \multicolumn{4}{c|}{V$_\mathrm{X}$} \\
\hline
\multicolumn{2}{|c|}{} & \multicolumn{4}{c|}{Missing Iodine} \\
\hline
\multicolumn{2}{|l|}{$\alpha$-CsPbI$_3$} & 3.37 & 96.8 & 0.51 & 42.8 \\
\multicolumn{2}{|l|}{strained CsPbI$_3$} & 23.4 & 130 & 0.26 & 0.95 \\
\hline
\multicolumn{2}{|c|}{} & \multicolumn{4}{c|}{Missing Bromine} \\  
\hline
\multicolumn{2}{|l|}{$\alpha$-CsPbBr$_3$} & 1.31 & 38.9 & 0.51 & 18.7 \\
\multicolumn{2}{|l|}{strained CsPbBr$_3$} & 6.62 & 51.8 & 0.31 & 0.89 \\
\hline
od-CsPbBr$_2$I & X$=$Br & 0.15 & 0.70 & 0.23 & 0.00 \\
& X$=$I & 19.1 & 238 & 0.38 & 13 \\
\hline
od-CsPbBrI$_2$ & X$=$Br & 0.11 & 8.77 & 0.66 & 47.6 \\
& X$=$I & 7.66 & 136 & 0.43 & 18.1 \\
\hline
ud-CsPbBr$_2$I & X$=$Br & 0.49 & 23 & 0.58 & 34.4 \\
& X$=$I & 0.04 & 82.8 & 1.17 & 2.06$\times10^{6}$ \\
\hline
ud-CsPbBrI$_2$ & X$=$Br & 0.11 & 8.77 & 0.66 & 47.6 \\
& X$=$I & 7.66 & 136 & 0.43 & 18.1 \\
\hline
\hline
\multicolumn{2}{|c|}{} & \multicolumn{4}{c|}{V$_\mathrm{Cs}$} \\  
\hline
\multicolumn{2}{|l|}{$\alpha$-CsPbI$_3$} & 0.04 & 454 & 1.41 & 5.93$\times10^{8}$ \\
\multicolumn{2}{|l|}{strained CsPbI$_3$} & 33.6 & 254 & 0.31 & 3.99 \\
\hline
\multicolumn{2}{|l|}{$\alpha$-CsPbBr$_3$} & 0.01 & 115 & 1.51 & 9.09$\times10^{8}$ \\
\multicolumn{2}{|l|}{strained CsPbBr$_3$} & 39.1 & 91.7 & 0.13 & 0.08 \\
\hline
\multicolumn{2}{|l|}{od-CsPbBr$_2$I}  & 4.61 & 80 & 0.43 & 10 \\
\multicolumn{2}{|l|}{od-CsPbBrI$_2$}  & 33.4 & 224 & 0.29 & 2.61 \\
\multicolumn{2}{|l|}{ud-CsPbBr$_2$I}  & 0.04 & 42.5 & 1.04 & 1.34$\times10^{5}$ \\
\multicolumn{2}{|l|}{ud-CsPbBrI$_2$}  & 17.6 & 141 & 0.31 & 2.56 \\
\hline

      \hline
    \end{tabular}
  \end{table*}

 \begin{table*}[h!]
\centering
\caption{Activation energies and pre-exponential factors from Neural Network Potentials, developed by \citeauthor{Tyagi2026}.\cite{Tyagi2025,Tyagi2026}. To convert the calculated pre-exponential factors $D_0$ from this work to the concentration independent pre-exponential factors of the defect $D_\mathrm{0,defect}$, we have divided them by the number of mobile halides $N_\mathrm{X}$ in the respective supercell.}
\label{tab:activation_energies_viren}
\begin{tabular}{|l l|l|l|l|l|}
  
\hline
\multicolumn{2}{|l|}{$D$ in [cm$^2\,$s$^{-1}$]}& $E_a$ [eV] & $D_\mathrm{0}$ $[\times 10^{-6}]$ & $N_\mathrm{X}$   & $D_\mathrm{0,defect},$ $[\times 10^{-3}]$ \\
\hline
\multicolumn{2}{|l}{I$_\mathrm{X}$} &  \multicolumn{4}{c|}{} \\
\hline

\multicolumn{2}{|l|}{$\alpha$-CsPbI$_3$ - this work} & 0.11 & 1.79 & 193 & 0.35 \\
\multicolumn{2}{|l|}{$\alpha$-CsPbI$_3$ - from Ref.~\citenum{Tyagi2025}} & 0.17 ± 0.02 &   & 649 & 0.80 ± 0.41\\

\hline
\multicolumn{2}{|l|}{$\alpha$-CsPbBr$_3$ - this work} & 0.09 & 1.07 &  193 & 0.21 \\
\multicolumn{2}{|l|}{$\alpha$-CsPbBr$_3$ - from Ref.~\citenum{Tyagi2026}} & 0.20 ± 0.02 &  & 649 & 1.54 ± 0.74\\
\hline
\hline
\multicolumn{2}{|l}{V$_\mathrm{X}$} & \multicolumn{4}{c|}{} \\  
\hline
\multicolumn{2}{|l|}{$\alpha$-CsPbI$_3$ - this work} & 0.51 &  42.8 & 191 & 8.17 \\
\multicolumn{2}{|l|}{$\alpha$-CsPbI$_3$ - from Ref.~\citenum{Tyagi2025}} & 0.26 ± 0.03 &  & 647 &  1.17 ± 0.91\\
\hline
\multicolumn{2}{|l|}{$\alpha$-CsPbBr$_3$ - this work} & 0.51 & 18.7 & 191 & 3.57 \\
\multicolumn{2}{|l|}{$\alpha$-CsPbBr$_3$ - from Ref.~\citenum{Tyagi2026}} & 0.26 ± 0.04 &  & 647 &  0.84 ± 0.81\\

      \hline
    \end{tabular}
  \end{table*}

\clearpage

\subsection{Forces}\label{ssec:forces}

\begin{figure*}
  \centering
  \includegraphics[]{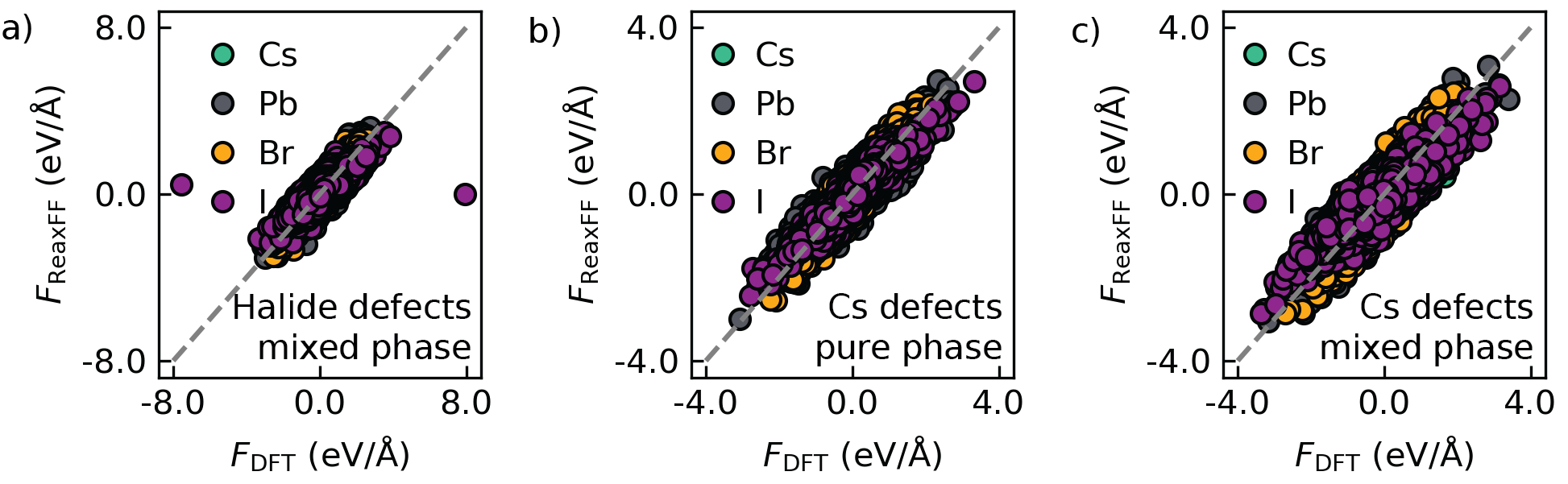}  
  \caption{Comparison between atomic forces calculated with ReaxFF and DFT for defect-containing perovskite snapshots at 700~K. Panel a) shows mixed-halide systems with halide defects, panel b) pure CsPbBr$_3$ and CsPbI$_3$ systems with Cs defects, and panel c) mixed-halide systems with Cs defects.}
  \label{fig:force_validation}
\end{figure*}

The validation of ReaxFF-calculated forces against DFT was performed by comparing forces for 21 snapshots, extracted in intervals of 500~ps from all molecular dynamics runs, containing a defect at 700~K. Snapshots were evaluated using VASP. DFT calculations employed the projector-augmented wave\cite{Kresse1999} method, treating Br (4s$^{2}$4p$^{5}$), Cs (5s$^{2}$5p$^{6}$6s$^{1}$), and Pb (6s$^{2}$6p$^{2}$) as valence electrons, with the PBE-GGA functional and D3-BJ van der Waals corrections\cite{Perdew1996,Grimme2011}. A $1\times1\times1$ Monkhorst-Pack k-grid\cite{Monkhorst1976} and 300 eV kinetic energy cutoff were used, with SCF convergence set to 10$^{-5}$ eV. Although ReaxFF has a polarization scheme to account for local charges, the DFT has been carried out with a neutral charge state, as the ReaxFF trainingset does not include charged defects.
Forces from both methods were directly compared to assess agreement. Figure~\ref{fig:force_validation} summarizes the parity between ReaxFF and DFT forces for the three defect classes considered here, while the detailed numerical results are displayed in Table~\ref{tab:forces_X_mixed}, \ref{tab:forces_Cs_pure} and \ref{tab:forces_Cs_mixed}. The NNPs, trained by \citeauthor{Tyagi2026} \cite{Tyagi2026} typically achieve a mean absolute error of 0.05~eV/\AA{} and a coefficient of determination of around 0.95. However, the ReaxFF model still provides reasonable force predictions, with overall mean absolute errors ranging from 0.129 to 0.159~eV/\AA{} and coefficients of determination between 0.638 and 0.817 across different elements and defect types. The largest force deviations are observed for Pb atoms, which is consistent with the known challenges in accurately modeling heavy elements in force fields. As the velocity errors scale with 1/m$_\mathrm{atom}$, this is less critical for the overall dynamics, and the ReaxFF model still captures the essential physics of defect migration in these systems. Overall, while NNPs may offer higher accuracy, ReaxFF provides a good balance between computational efficiency and force prediction quality for the defect-diffusion studies presented here.

 \begin{table*}[h!]
\centering
\caption{Per-element force prediction performance of the ReaxFF force field\cite{Pols2024} for 336 time frames of mixed-halide perovskites containing a halide defect, benchmarked against DFT reference data. Reported are the number of force components ($N_\mathrm{Forces}$), the mean absolute error ($F_\mathrm{MAE}$, eV/\AA{}), and the coefficient of determination ($F_{\mathrm{R}^2}$).}
\label{tab:forces_X_mixed}
\begin{tabular}{|l|l|l|l|}
  \hline
  Element & $N_\mathrm{Forces}$ & $F_\mathrm{MAE}$ [eV/\AA{}] & $F_\mathrm{R^2}$ \\
  \hline
Cs &64512 & 0.102 & 0.707 \\
Pb &64512 & 0.216 & 0.748 \\
Br &96768 & 0.152 & 0.638 \\
I &96768 & 0.165 & 0.709 \\
All &322560 & 0.159 & 0.709 \\

      \hline
    \end{tabular}
  \end{table*}

 \begin{table*}[h!]
\centering
\caption{Per-element force prediction performance of the ReaxFF force field\cite{Pols2024} for 42 time frames each of pure CsPbBr$_3$ and CsPbI$_3$ perovskites containing a Cs defect, benchmarked against DFT reference data. Reported are the number of force components ($N_\mathrm{Forces}$), the mean absolute error ($F_\mathrm{MAE}$, eV/\AA{}), and the coefficient of determination ($F_{\mathrm{R}^2}$).}
\label{tab:forces_Cs_pure}
\begin{tabular}{|l|l|l|l|}
  \hline
  Element & $N_\mathrm{Forces}$ & $F_\mathrm{MAE}$ [eV/\AA{}] & $F_\mathrm{R^2}$ \\
  \hline
Cs & 16128 & 0.092 & 0.730 \\
Pb & 16128 & 0.189 & 0.798 \\
Br & 24192 & 0.124 & 0.759 \\
I & 24192 & 0.117 & 0.817 \\
All & 80640 & 0.129 & 0.789 \\

      \hline
    \end{tabular}
  \end{table*}

 \begin{table*}[h!]
\centering
\caption{Per-element force prediction performance of the ReaxFF force field\cite{Pols2024} for 168 time frames of mixed-halide perovskites containing a cs defect, benchmarked against DFT reference data. Reported are the number of force components ($N_\mathrm{Forces}$), the mean absolute error ($F_\mathrm{MAE}$, eV/\AA{}), and the coefficient of determination ($F_{\mathrm{R}^2}$).}
\label{tab:forces_Cs_mixed}
\begin{tabular}{|l|l|l|l|}
  \hline
  Element & $N_\mathrm{Forces}$ & $F_\mathrm{MAE}$ [eV/\AA{}] & $F_\mathrm{R^2}$ \\
  \hline
  Cs &32256 & 0.104 & 0.691 \\
  Pb &32256 & 0.215 & 0.754 \\
  Br &48384 & 0.151 & 0.644 \\
  I &48384 & 0.162 & 0.719 \\
  All &161280 & 0.158 & 0.714 \\
      \hline
    \end{tabular}
  \end{table*}

  \clearpage

\subsection{Tilt Geometries}
\begin{figure*}[h!]
    \centering
    \includegraphics[]{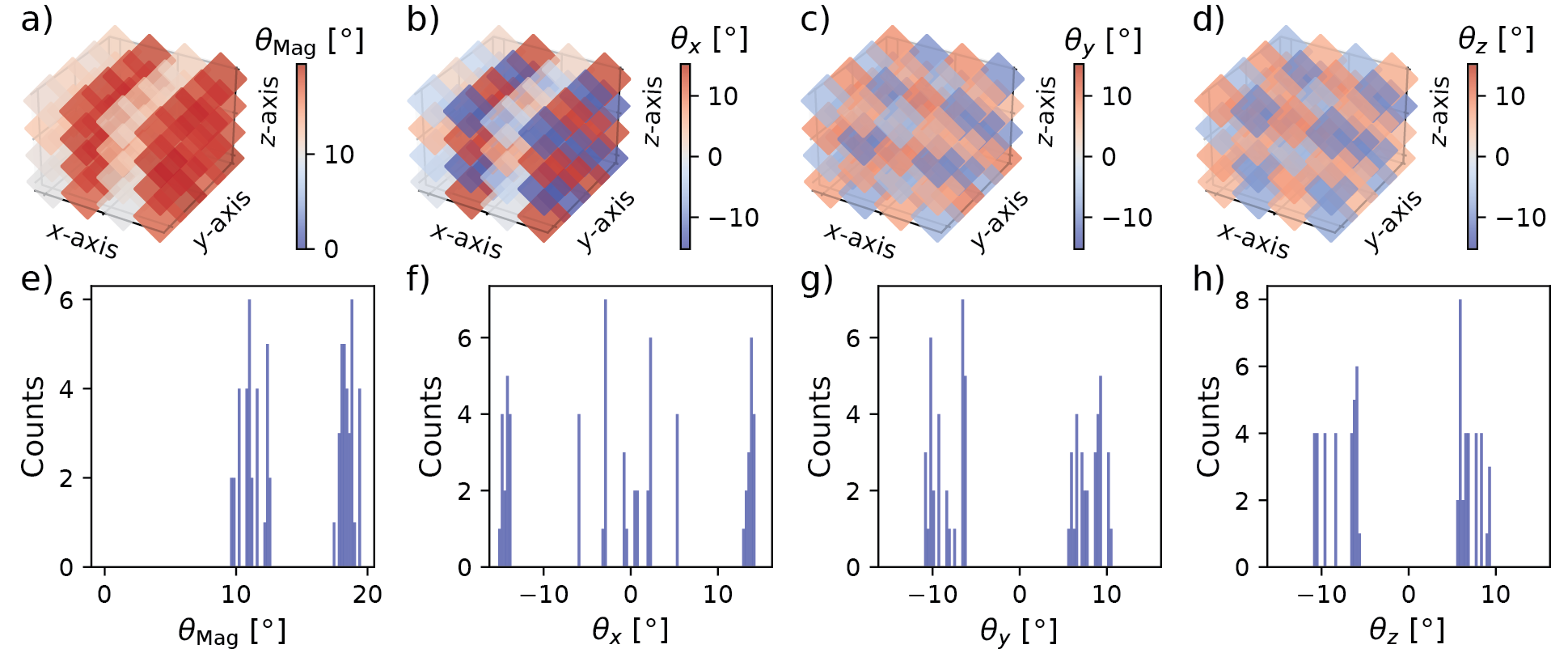}
\caption{Of the DFT optimized configurations of layered CsPbBr$_2$I$_{1}$ perovskites: a) Rotation magnitude of the PbX$_6$ octahedral tilts and b)-d) time-averaged tilt angles with respect to the cartesian axes ($\theta_x,\theta_y,\theta_z$) in  perovskites. The color scale indicates the angle (in degrees) for each octahedron. Panel e)-h) are the histograms, referring to the respective pattern above, with 100 bins. For the I octahedrons we observe $\theta_\mathrm{Mag}\approx 18.2^{\circ}$ and $\theta_x\approx \pm 14.9^{\circ}$. For the Br octahedrons we observe $\theta_\mathrm{Mag}\approx 11.0^{\circ}$ and $\theta_x\approx \pm 0^{\circ}$.}
    \label{fig:3-1_tilt_pattern_GO}
  \end{figure*}

\begin{figure*}[h!]
    \centering
    \includegraphics[]{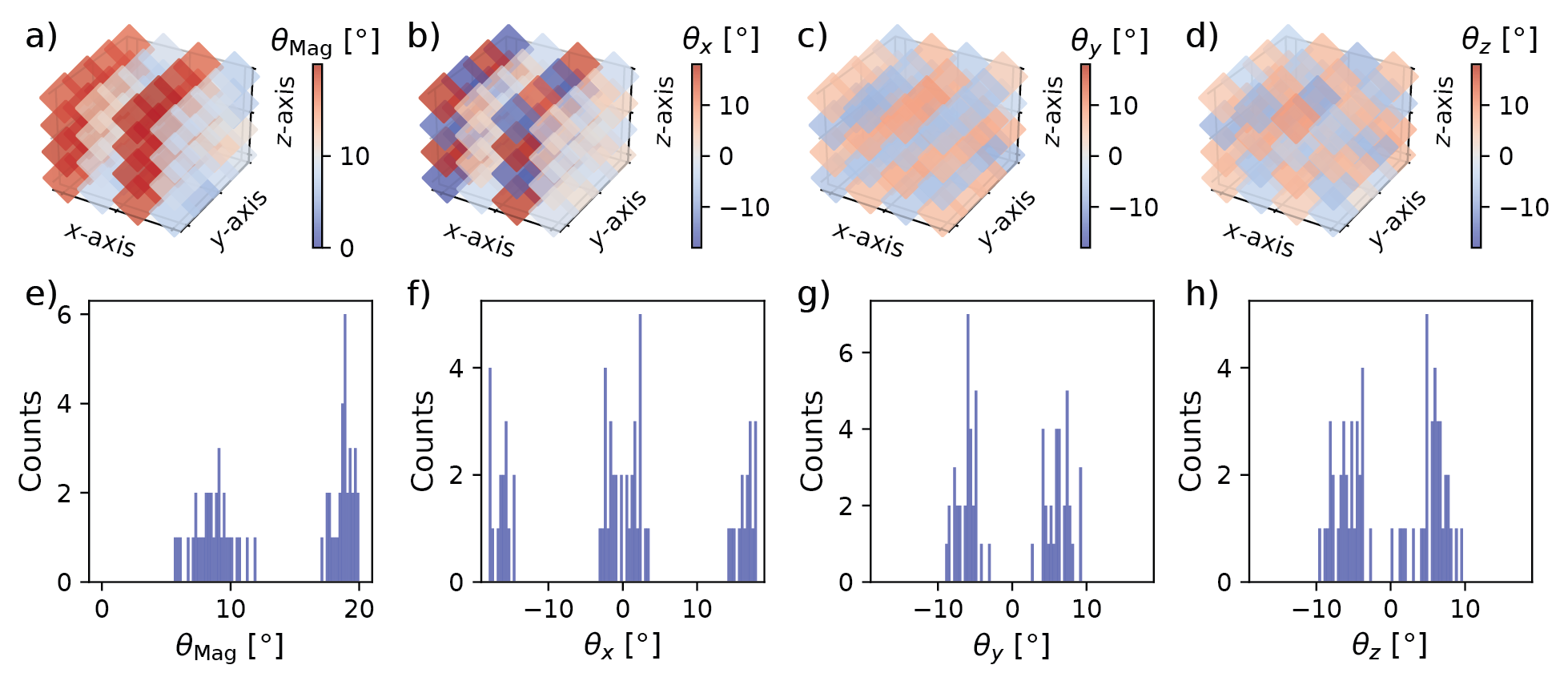}
\caption{Of the DFT optimized configurations of layered CsPbBrI$_2$ perovskites: a) Time-averaged rotation magnitude of the PbX$_6$ octahedral tilts and b)-d) time-averaged tilt angles with respect to the cartesian axes ($\theta_x,\theta_y,\theta_z$) in  perovskites. The color scale indicates the angle (in degrees) for each octahedron. Panel e)-h) are the histograms, referring to the respective pattern above, with 100 bins. For the I octahedrons we observe $\theta_\mathrm{Mag}\approx 18.9^{\circ}$ and $\theta_x\approx \pm 15.5^{\circ}$. For the Br octahedrons we observe $\theta_\mathrm{Mag}\approx 8.7^{\circ}$ and $\theta_x\approx \pm 0^{\circ}$.}
    \label{fig:1-3_tilt_pattern_GO}
  \end{figure*}

\begin{figure*}[h!]
    \centering
    \includegraphics[]{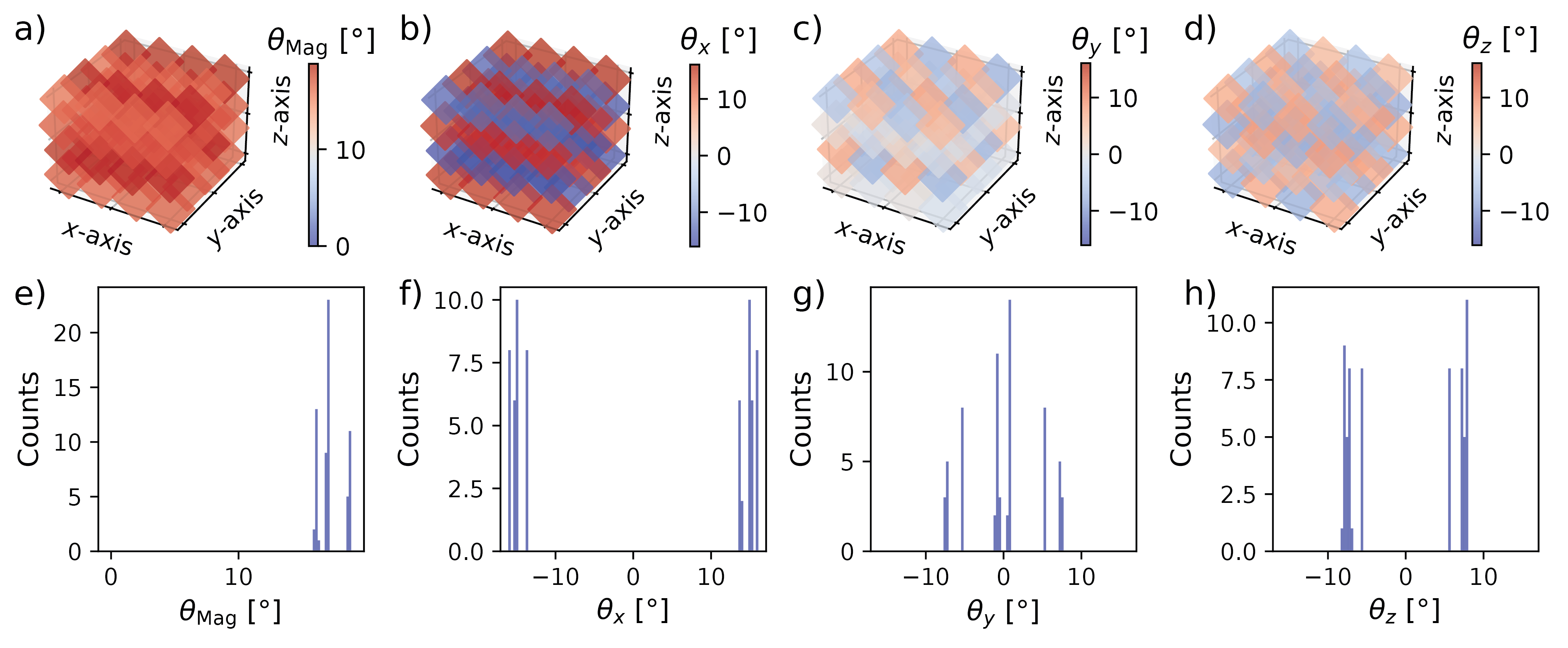}
  \caption{Of the DFT optimized configurations of CsPbI$_3$ under biaxial compression to achieve similar tilt magnitudes as in layered CsPbBr$_2$I$_{1}$. a) Time-averaged rotation magnitude of the PbX$_6$ octahedral tilts and b)-d) their components along the axes. The color scale indicates the angle (in degrees) for each octahedron. Panel e)-h) are the histograms, referring to the respective pattern above, with 100 bins. We observe $\theta_\mathrm{Mag}\approx 17.1^{\circ}$, while $\theta_x\approx \pm 15.0^{\circ}$}
    \label{fig:strained_cspbi3_tilt_pattern_GO}
    \end{figure*}

To assess the structural accuracy of tilts in ReaxFF, we compared its geometry-optimized configurations against those obtained from DFT for $4\times4\times4$ supercells of CsPbBr$_3$, CsPbI$_3$, and CsPbBr$_x$I$_{3-x}$. Geometry optimizations in VASP-DFT were performed by relaxing ionic positions of a single time step (with fixed cell shape and volume) using a conjugate gradient algorithm and a maximum of 1,000 ionic steps. Other than that, the same parameters as for the force evaluation in Subsection~\ref{ssec:forces} were applied. The tilting patterns of the PbX$_6$ octahedra in mixed halide perovskites in Fig.~\ref{fig:3-1_tilt_pattern_GO}, \ref{fig:1-3_tilt_pattern_GO} and \ref{fig:strained_cspbi3_tilt_pattern} are in excellent agreement with the time-averaged tilts in SI Note:~\ref{sec:tilts}. We note, that the compressed CsPbI$_3$ structure in Fig.~\ref{fig:strained_cspbi3_tilt_pattern_GO}, shows considerably larger tilts than the ones predicted by ReaxFF. However, this strengthes our conclusions, as higher levels of tilts are expected to further suppress the diffusion of Cs defects in the layered perovskites and might even influence halide migration. It is to note that the angular components along $y$- and $z$-axis appear to be non-zero, as this is only a single time frame, and not averaged over a full simulation of multiple nanoseconds.

    \clearpage

\subsection{Defect Geometries}

\subsubsection{Cesium Defects}
To assess the structural accuracy of cesium defects in ReaxFF, we compared its geometry-optimized configurations (in AMS2024, with quality "Good") against those obtained from DFT for $4\times4\times4$ supercells of CsPbBr$_3$, CsPbI$_3$, and CsPbBr$_x$I$_{3-x}$. Geometry optimizations in VASP-DFT were performed by relaxing ionic positions (with fixed cell shape and volume) using a conjugate gradient algorithm and a maximum of 1,000 ionic steps. The evaluated frames are spaced by 1~ns. Other than that, the same parameters as for the force evaluation in Subsection~\ref{ssec:forces} were applied. The same supercells were optimized using ReaxFF, and the resulting structures were directly compared to the DFT-optimized geometries. For comparison, we took the 8 surroundig Pb atoms of the Cs defect and calculated the volume of the cuboid they span. The cubic root of this volume is reported as the average Pb-Pb distance $d_\mathrm{Pb}$ in Tables~\ref{tab:Cs_Vac_GO} and \ref{tab:Cs_Int_GO}. Furthermore, we used it to calculate the volumetric strain $\epsilon$ of the defect, defined as $\epsilon = (4\times4\times4\times \Delta V / V_\mathrm{supercell}) \times 100$. A positive value indicates an expansion, while a negative value indicates a contraction. Both measures are well captured by ReaxFF, especially for the mixed-halide systems. For I$_\mathrm{Cs}$, we also report the average distance of the interstital Cs atoms. Those are generally overestimated by around 0.4~\AA{}.

\begin{table}[h!]
  \caption{Structural comparison of Cs vacancy defects between ReaxFF and DFT: Pb-Pb distances and volumetric strain $\epsilon$.}
\label{tab:Cs_Vac_GO}
\begin{tabular}{|l|c|c|c|c|c|}
\hline
  Structure & N$_\mathrm{F}$ & $d_\mathrm{Pb,DFT}$[\AA{}] & $d_\mathrm{Pb,ReaxFF}$[\AA{}]& $\epsilon_{\mathrm{DFT}}$[\%] & $\epsilon_{\mathrm{ReaxFF}}$[\%] \\
\hline
\setlength{\tabcolsep}{3pt} 
od-CsPbBrI$_2$ & 10 & 6.23$\pm$0.01 & 6.29$\pm$0.01 & 0.65 & 3.60 \\
ud-CsPbBrI$_2$ & 9 & 6.41$\pm$0.02 & 6.56$\pm$0.02 & 8.58 & 16.43 \\
od-CsPbBr$_2$I & 8 & 6.17$\pm$0.01 & 6.20$\pm$0.01 & 4.89 & 5.96 \\
ud-CsPbBr$_2$I & 9 & 6.20$\pm$0.02 & 6.18$\pm$0.01 & 9.60 & 8.36 \\
$\alpha$-CsPbBr$_3$ & 10 & 5.99$\pm$0.01 & 6.14$\pm$0.01 & 1.88 & 9.63 \\
$\alpha$-CsPbI$_3$ & 7 & 6.44$\pm$0.01 & 6.53$\pm$0.01 & 2.39 & 7.07 \\
\hline
\end{tabular}

\end{table}

\begin{table}[h!]
  \caption{Structural comparison of Cs interstitial defects between ReaxFF and DFT: Pb-Pb and Cs-Cs distances and volumetric strain $\epsilon$.}
  \label{tab:Cs_Int_GO}

\begin{tabular}{|l|c|c|c|c|c|c|c|}
\hline
\setlength{\tabcolsep}{-15pt} 

Structure & N$_\mathrm{F}$ & $d_\mathrm{Pb,DFT}$[\AA{}] & $d_\mathrm{Pb,ReaxFF}$[\AA{}]& $\epsilon_{\mathrm{DFT}}$[\%] & $\epsilon_{\mathrm{ReaxFF}}$[\%] & $d_\mathrm{Cs,DFT}$[\AA{}] & $d_\mathrm{Cs,ReaxFF}$[\AA{}] \\
\hline

od-CsPbBrI$_2$ & 5 & 6.44$\pm$0.03 & 6.47$\pm$0.03 & 11.23 & 12.85 & 3.81$\pm$0.02 & 4.21$\pm$0.03 \\
ud-CsPbBrI$_2$ & 11 & 6.55$\pm$0.03 & 6.72$\pm$0.05 & 16.18 & 25.24 & 3.87$\pm$0.03 & 4.21$\pm$0.03 \\
od-CsPbBr$_2$I & 7 & 6.29$\pm$0.02 & 6.26$\pm$0.01 & 10.95 & 9.51 & 3.77$\pm$0.01 & 4.09$\pm$0.01 \\
ud-CsPbBr$_2$I & 7 & 6.27$\pm$0.02 & 6.16$\pm$0.01 & 13.13 & 7.48 & 3.70$\pm$0.06 & 4.12$\pm$0.04 \\
$\alpha$-CsPbBr$_3$ & 8 & 6.10$\pm$0.01 & 6.14$\pm$0.02 & 7.14 & 9.48 & 3.65$\pm$0.01 & 4.15$\pm$0.03 \\
$\alpha$-CsPbI$_3$ & 5 & 6.56$\pm$0.00 & 6.54$\pm$0.02 & 8.29 & 7.26 & 3.87$\pm$0.01 & 4.20$\pm$0.03 \\
\hline

\end{tabular}

\end{table}

\clearpage

  \bibliography{Paper_Layers}